\documentclass[aps,prb,twocolumn,groupedaddress,showpacs,floatfix,eqsecnum]{revtex4}
\usepackage{graphics,graphicx}
\usepackage{color}
\usepackage{wrapfig}
\usepackage{enumerate}
\usepackage{amssymb,amsmath}
\usepackage{bookmath}
\usepackage{bm}

\newcommand{\req}[1]{Eq.~(\ref{#1})}
\newcommand{\reqs}[1]{Eqs.~(\ref{#1})}
\newcommand{\rfig}[1]{Fig.~\ref{#1}}
\newcommand{\pp}{ {{+}{+}} }
\newcommand{\es}{ {{+}{-}} }
\newcommand{\whSigma}{{\widehat\Sigma}}
\newcommand{\delt}{{\delta t}}

\begin{document}
\unitlength = 1mm
\title{Superconductivity and spin-density-waves in multi-band metals}
\author{A.~B.~Vorontsov$^1$, M.~G.~Vavilov$^2$ and A.~V.~Chubukov$^2$}
\affiliation{
$^1$~Department of Physics, Montana State University, Bozeman, Montana, 59717, USA\\
$^2$~Department of Physics, University of Wisconsin, Madison, Wisconsin 53706, USA
}

\date{\today}

\begin{abstract}
We present a detailed description of two-band quasi-2D metals with $s$-wave 
superconducting (SC) and antiferromagnetic spin-density wave (SDW)
correlations. We present  a general approach and use it to
investigate the influence of
the difference between the shapes and the areas of the two
Fermi surfaces on the phase diagram.
In particular, we determine 
the conditions for the co-existence of SC and SDW orders at different
temperatures and dopings. We argue that a conventional $s$-wave SC order
co-exists with SDW order only  at very low $T$ and in a very
tiny range of parameters. 
An extended $s$-wave superconductivity, 
for which SC gap changes sign between the two bands, 
co-exists with antiferromagnetic SDW over a much wider range of parameters 
and temperatures, but even for this SC order the regions of SDW and SC can
still be separated by a first order transition. We show that the co-existence
range  becomes larger if SDW order is incommensurate. We apply our results to
iron-based pnictide materials, in some of which co-existence of SDW and SC
orders has been detected. 
\end{abstract}
\maketitle

\section{Introduction}

Discovery of new magnetically-active superconductors, 
iron pnictides, based on 
FeAs\cite{Kamihara08,Rotter08} or Fe(Se,S,Te)\cite{Hsu08FeSe,Subedi08FeSe} 
has further invigorated the on-going discussions about 
co-existence of different ordered electronic states in metals.
\cite{Bulaevskii85,Onuki04magsc,Steglich07} 
In itinerant electrons systems,
the interactions that lead to 
formation of superconducting (SC) and magnetic spin-density-wave (SDW) 
orders, ``pull'' and ``push'' the same particles, and as a result,
influence each other.
In particular, two orders may 
support each other and lead to
homogeneous local co-existence of SC and SDW states; 
or one of them 
may completely suppress the other order, resulting in a state 
with spatially separated
regions of ``pure'' SDW or SC orders.
The transitions between various states may also be either 
continuous (second order) or abrupt (first order). 
The outcome of this interplay depends critically on 
a number of parameters: 
 properties of the
interactions, such as symmetry of SC pairing, 
their relative strengths, 
and also on properties the Fermi surface (FS), such as
its shape or the density of electronic states. 

In pnictides this parameter space is vast. 
First, these are multi-band materials, with 
two hole pockets in the center, $(0,0)$, 
and two electron pockets 
near $(\pm\pi, 0)$ and $(0, \pm\pi)$ points
of the unfolded Brillouin zone (BZ) 
(one Fe atom per unit cell). The shapes of 
quasi two-dimensional electron 
pockets are quite distinct in different materials, 
ranging from simple circle-like types in LaOFeP\cite{Lu08elstrct1111,Lu09arpes1111}, 
to cross-like electronic FS in LaOFeAs,\cite{Lu09arpes1111} 
to ellipses in BaFe$_2$As$_2$\cite{Liu08fs122,Yi09elstrct122} 
and even more complex propeller-like structures in 
(Ba,K)Fe$_2$As$_2$\cite{Zabolotnyy09elord} 
(for a descending point of view on this see Ref.~\onlinecite{Yi09elstrct122}).
Hole pockets are near-circular, but different 
hole pockets in the same material usually have different sizes.

Second, multiple FSs also create a number of different
possibilities\cite{Chubukov08,Chubukov09rg} for electron ordering 
in the form of SDW, charge density wave (CDW) states,
and various superconducting states. 
The SC states include
1) the conventional 
$s^\pp$-wave state that has $s$-wave symmetry 
in the BZ and gaps of the same sign on electron and hole FSs; 
2) the extended $s^\es$-state that looks as $s$-wave from a symmetry 
point of view but has opposite signs 
of the gaps on pockets at $(0,0)$ and $(\pm \pi,0)$,
\cite{Mazin08splus,Kuroki08,Barzykin08,Mazin09review}
and 3) several SC states with  the nodes in the SC gap, 
of both $s$-wave and $d$-wave symmetry.\cite{SeoBernevig08,Maier09gap,Graser09,Chubukov09nodes,Thomale09nodes} 

As a result of this complex 
environment, the interplay of magnetic 
and superconducting orders also 
shows some degree of variations. 
Most of parent compounds of
iron pnictides are magnetically ordered. Upon doping, magnetism
eventually yields to superconductivity, but 
how this transformation occurs 
varies significantly 
between different Fe-pnictides. 
A first-order transition between SC and SDW
orders has been reported for (La,Sm)O$_{1-x}$F$_x$FeAs.\cite{Luetkens09,Sanna09} 
 On the other hand, 
 in electron-doped Ba(Fe$_{1-x}$Co$_x$)$_2$As$_2$ 
 recent nuclear magnetic resonance (NMR),~\cite{Laplace09,Julien09} specific heat,
susceptibility, Hall coefficient,~\cite{Chu09,Rotter08} and neutron scattering
experiments~\cite{Fernandes09} 
indicate that SDW and SC phases coexist 
locally over some doping range. 
In the same 122 family, experiments on hole-doped Ba$_{1-x}$K$_x$(FeAs)$_2$ 
disagree with each other and indicate both co-existence\cite{Chen09cxK122,Rotter09} and 
incompatibility\cite{Julien09,Park09K122,Goko09sdwsc122} of two orders. 
Isovalently doped 122 material BaFe$_2$(As$_{1-x}$P$_x$)$_2$ shows 
the region of coexistence. 
\cite{Shishido09fsevol,Kasahara09evolut,Hashimoto09gap}

The goal of the present work is to understand how the system evolves from an
SDW antiferromagnet to an 
 $s^{++}/s^{+-}$-wave  superconductor,
and how this evolution depends on the shape of the FS, the
strengths of the interactions, and the 
structure of the SC order.
For this we derive and solve a
set of coupled non-linear BCS-type equations for SC and SDW order
parameters and compare values of the free energy for possible phases.

We report several results. 
First, we find that there is much more inclination for co-existence 
between $s^{\es}$ and SDW orders 
than between the same-sign $s^\pp$-wave state and SDW. 
In the latter case, co-existence is only possible at very low $T$ and in a very
tiny range of parameters.
Second, the co-existence region generally grows with increased 
strength of SDW coupling relative to superconducting interaction. 
That the co-existence is only possible when SDW transition comes first 
has been noticed some time ago,\cite{Machida81sdwsc,Machida81b} and our results
agree with these findings. 
Third, when SDW order is commensurate, the co-existence is only possible when
the following 
two conditions are met simultaneously:  hole and electron FSs have different
$k_F$ (cross-section areas) {\it and} 
different shapes, (e.g., hole pockets are circles and electron pockets are ellipses). 
Even then, SDW and SC orders co-exist only in a limited range of
parameters and temperatures, see Sections \ref{sec:numer} and \ref{sec:GL},  
Figs.~\ref{fig:pdSPzeroT} and \ref{fig:GL} below. 
When SDW order is incommensurate, the difference in $k_F$ is 
a sufficient condition,
but again, the two orders co-exist in a limited range of
parameters/temperatures (\rfig{fig:cx35}).   

We also analyze in some detail the interplay between the co-existence and
the presence of the Fermi surface (i.e. gapless excitations) 
in the SDW state.  The ``conventional''
logic states that superconductivity and magnetism compete for the Fermi surface
and co-exist if SDW order still leaves a modified Fermi surface on which SC
order can form. We find that the situation is more complex and the mere
 presence of absence of a modified Fermi surface is not the key reason for
co-existence. We show that 
a more important reason is the 
effective ``attraction''
between SDW and SC order parameters,
when the development of one order favors a gradual 
formation of the other order.
Specifically, we show that: 

\vspace{1mm}
\begin{minipage}{0.95\linewidth}
 $\bullet$ 
near the point where the transitions from the normal (N) state into SC and SDW
states cross, SC can develop either via the co-existence phase or via a direct
first order transition between pure SDW and SC states, In this range, the SDW
order parameter is small and SDW state is definitely a metal, \rfig{fig:GL};

\vspace{1mm}
$\bullet$ 
at low $T$, the co-existence phase may develop even when SDW state has no Fermi
surface (not counting bands which do not participate in SDW). In this situation
there is no Fermi surface for a conventional development of the SC order, but
the system still can lower the energy by developing both orders, if there is an
``attraction'' between them.  This is the case for $s^{+-}$ superconductivity and
comparable strength of SDW and SC couplings, Figs.~\ref{fig:tsc2pd},
\ref{fig:pdSPzeroT}(a);

\vspace{1mm}
$\bullet$ 
the SDW phase at low $T$ can be a metal with rather large Fermi surfaces, yet
SC order does not develop. This is the case when SC order is $s^{++}$,
\rfig{fig:SSzeroT}. 
\end{minipage}
\vspace{1mm}

The close connection between the co-existence of the
two states and the symmetry of the SC state 
has been discussed earlier in the context 
of single-band heavy-fermion materials.\cite{MachidaKato87} 
This connection 
gives a possibility to obtain information about the pure states 
(e.g., about the structure of the SC gap) from experimental investigations of the SC - SDW
interplay,  as it has been recently suggested.\cite{Fernandes09,Parker09} 

The structure of the paper is as follows. In the next section we 
define the model and derive generic 
equations for the SDW and SC order parameters and an expression for the free energy. Then we 
simplify these formulas for the case of a small splitting between hole and
electron FSs and utilize them 
in Secs.~\ref{sec:pureSDW} through ~\ref{sec:GL}.
In Sec.~\ref{sec:pureSDW} we focus on a pure SDW state, 
with special attention given to the interplay
between ellipticity of the FS and the incommensuration of the SDW order. 
In the next two sections we discuss possible co-existence of SDW and SC states: in
Sec.~\ref{sec:numer} we present numerical results obtained in a wide range of
temperatures and dopings, and in Sec.~\ref{sec:GL} we corroborate this with the
analytical consideration in the vicinity of the crossing point 
of  SC and SDW transitions, and at $T=0$.
In Sec.~\ref{sec:partSDW} we model
the case when the splitting between the two FSs is not small. 
We present our conclusions in Sec.~\ref{sec:concls}.  
Some of the results reported in this work have been presented in shorter
publications.\cite{Vavilov09gl,Vorontsov09magsc}

\begin{figure}[t]
\centerline{\includegraphics[width=\linewidth]{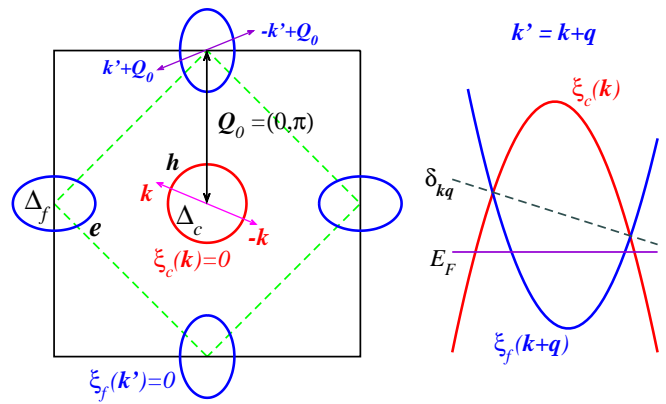}}
\caption{(Color online) 
Left:  electronic structure of the two band model considered in this paper, 
in the unfolded Brillouin zone.   
The hole FS is in the center, with SC order parameter $\Delta_c$, and 
the electron FSs are  at $(0,\pi)$ and $(\pi,0)$,  
with SC order parameter  $\Delta_f$.  The magnetic order with momentum 
$\vQ_0 = (0,\pi)$ hybridize  hole and electron FSs separated by $\vQ_0$, 
but leaves FSs at $(\pm \pi,0)$ intact.  
 Right: by doping or pressure one may adjust 
the size and shape of hole and electron bands,
and also SDW order parameter can be incommensurate, with momentum $\vQ_0 + \vq$. 
These effects are described by FS detuning parameter, 
$\delta_{\vk \vq} = [\xi_f(\vk+\vq)+\xi_c(\vk)]/2$.
}
\label{fig:FS}
\end{figure}

\section{Model and analytical reasoning}
\label{sec:model}

\subsection{General formulation} 

Since the basic properties of the SC and magnetic SDW
interactions and their interplay should not depend on the 
number of bands significantly, we consider a basic model of 
one hole and one electron bands.  
For pnictides this means that we neglect the double degeneracy 
of hole and electron states at the center and the 
corners of 
the Brillouin zone, which does not seem to be essential 
for superconducting 
\cite{Maier09gap,Chubukov09nodes,Thomale09nodes,Cvetkovic09,WangLee09,Platt09,Raghu08}
or magnetic order.
\cite{Lorenzana08,Brydon09,Johannes09,Mazin09,Eremin09}

The basic model is illustrated in \rfig{fig:FS}. 
Electronic structure contains two families of fermions,
near one hole and one electron FSs of small and near-equal sizes. 
Such two-band structure yields the experimentally observed stripe $(\pi,0)$ 
or $(0,\pi)$ magnetic order which in itinerant scenario appears, at least
partly, due to nesting between one hole and one electron bands, separated by
momentum $(\pi,0)$ or $(0,\pi)$. 
Other hole and electron bands do not participate in the SDW order.
 We assume that SC also primarily resides on the same two 
FSs, at least close to the boundary of the SDW phase. The SC order parameter 
on the other two bands is not zero, but is smaller.  Once doping increases and
the system moves away from SDW boundary, we expect that the magnitudes of the 
SC order parameter on the two electron bands should become closer to each other.

The basic Hamiltonian includes the free fermion part $\cH_0$,  
and the fermion-fermion interactions 
in superconducting and magnetic SDW channels, 
\be
\label{eq:cHfull}
\cH = \cH_0 + \cH_\Delta + \cH_m \,.
\ee
The free fermion part of the Hamiltonian is
\be
\cH_0 = \sum_{\vk} \xi_c(\vk) c^\dag_{\vk\alpha} c_{\vk\alpha} 
+\sum_{\vk'} \xi_f(\vk') f^\dag_{\vk' \alpha} f_{\vk' \alpha} \,,
\ee
where creation/annihilation $c^\dag, c$-operators correspond to fermions 
near the central hole pocket $(0,0)$, and $f$-operators describe fermions 
near the electron pocket at $\vQ_0 = (0, \pi)$ and the fermion dispersions 
near the pockets are
\be
\xi_c (\vk) = \mu_c - \frac{\vk^2}{2m_c}, ~~ \xi_f (\vk ) = 
\frac{k_x^2}{2m_{fx}} + \frac{k_y^2}{2m_{fy}} -\mu_f 
\label{eq:xis}
\ee
The momenta $\vk$ are measured from the center of the BZ, 
and $\vk'$ are deviations from $\vQ_0$.  
We assume an inversion symmetry, $\xi_{c,f}(-\vk)=\xi_{c,f}(\vk)$. 

The pairing interaction consists of many 
different pair scattering terms, but the most important one 
is the pair hopping between the hole 
and electron pockets,\cite{Chubukov08,Chubukov09nodes} 
\bea
\label{eq:scH}
\cH_\Delta = \onehalf \sum_{\vk,\vp} V^{cf}_{\alpha\beta\beta'\alpha'}(\vk,\vp) 
\left( c^\dag_{\vk\alpha} c^\dag_{-\vk\beta} f_{-\vp\beta'} f_{\vp\alpha'} 
\right.  \\ \left. \nonumber 
+ f^\dag_{\vk \alpha} f^\dag_{-\vk \beta} c_{-\vp \beta'} c_{\vp \alpha'} \right) \,,
\eea
For definiteness, we consider 
SC interaction only in the singlet channel, i.e.  
\be
V^{cf}_{\alpha\beta\beta'\alpha'}(\vk,\vp) = V^{sc}_{\vk,\vp} \;  
(i\sigma^y)_{\alpha\beta} (i\sigma^y)^\dag\,_{\beta'\alpha'} \,.
\ee

The magnetic interaction between fermions is 
\bea
 \label{eq:sdwH}
\cH_m &=& - \onefourth \sum_{\vp'-\vp = \vk'-\vk} \; 
V^{sdw}_{\alpha\beta\beta'\alpha'} (\vp'\vp;\vk,\vk')  \times
 \\  \nonumber 
&& \times \left( 
f^\dag_{\vp' \alpha} c_{\vp\beta} \; c^\dag_{\vk \beta'} f_{\vk' \alpha'}
+f^\dag_{-\vp' \alpha} c_{-\vp\beta} \; c^\dag_{-\vk \beta'} f_{-\vk' \alpha'} 
\right) \,,
\eea
where we symmetrized the expression with respect to particle hopping between 
$(0,0)-(0,\pi)$ and $(0,0)-(0,-\pi)$  pockets for later convenience. 
We take 
the interaction matrix in a simple form, 
\be
V^{sdw}_{\alpha\beta\beta'\alpha'} (\vp'\vp;\vk\vk') = V^{sdw}_{\vp'\vp;\vk\vk'} \; 
\vsigma_{\alpha\beta} \cdot \vsigma^\dag_{\beta'\alpha'} \,, 
\ee
with a constant $V^{sdw}_{\vp'\vp;\vk\vk'} = V^{sdw}$. 

The evolution of the interaction couplings with energy was 
considered in Ref.~\onlinecite{Chubukov08}. Here we assume that the interactions for
low-energy excitations can be represented in terms of fermion couplings to
order parameter fields in the SC and SDW channels. 
In the spirit of BCS-type approach, we introduce the SC order parameters  
 \begin{subequations}\label{eqs:OP}
\begin{align}
\Delta_c(\vk)_{\alpha\beta} & = (i\sigma^y)_{\alpha \beta} \; 
\sum_{\vp} V^{sc}_{\vk,\vp} \;  (i\sigma^y)^\dag_{\beta'\alpha'}
\langle f_{-\vp\beta'} f_{\vp\alpha'} \rangle \,,
\\
\Delta_f(\vk)_{\alpha\beta} &  = (i\sigma^y)_{\alpha \beta} \; 
\sum_{\vp} V^{sc}_{\vk,\vp} \;  (i\sigma^y)^\dag_{\beta'\alpha'}
\langle c_{-\vp \beta'} c_{\vp \alpha'} \rangle \,,
\end{align}
and the SDW order parameter directed along $\hat\vm$.
We assume that SDW order parameter has a single ordering 
momentum  $\vQ = \vQ_0 + \vq$, in which case it is fully specified by 
$(m_\vq)_{\alpha\beta}= (\vm_\vq \vsigma)_{\alpha\beta} 
= m_\vq (\hat\vm \vsigma)_{\alpha\beta}$, where  
\begin{equation}\begin{split}
(m_\vq)_{\alpha\beta} & = - V^{sdw} 
\frac{1}{2} \sum_{\vp} \vsigma_{\alpha\beta}\cdot \vsigma^\dag_{\beta'\alpha'} 
\langle c^\dag_{\vp \beta'} f_{\vp+\vq \alpha'} \rangle 
\\
& = - V^{sdw}  
\frac{1}{2} \sum_{\vp} \vsigma_{\alpha\beta}\cdot \vsigma^\dag_{\beta'\alpha'} 
\langle f^\dag_{-\vp-\vq \beta'} c_{-\vp \alpha'} \rangle  \,.
\end{split}\end{equation}
\end{subequations}
Since 
$\langle c^\dag_{\vp \alpha} f_{\vp+\vq \beta} \rangle 
\sim (\vm_\vq \vsigma)_{\alpha\beta}$,  
the corresponding electronic magnetization, 
\be 
\vm(\vR) = \sum_\vp  \vsigma_{\alpha\beta} \left[ 
\langle c^\dag_{\vp \alpha} f_{\vp+\vq \beta} \rangle e^{i\vQ\vR} 
+ \langle f^\dag_{\vp+\vq \alpha} c_{\vp \beta} \rangle e^{-i\vQ\vR} \right]
\nonumber 
\ee
is $\vm_\vq \cos\vQ\vR$ for  real $\vm_\vq$ and is $\vm'_\vq \cos\vQ\vR -
\vm''_\vq \sin\vQ\vR$ for a complex $\vm_\vq = \vm^\prime_\vq + i \vm^{\prime\prime}_\vq$. 
In principle, SDW order parameter may contain several components 
with different ${\bf q}$, which 
could give rise to domain-like structures 
of $\vm(\vR)$. For recent studies in this direction 
see Ref.~\onlinecite{Gorkov10_solitonSDW}.  
We  perform the analysis of the co-existence between SC order and SDW order
with a single ${\bf q}$. 
A more general form of the SDW order should not qualitatively change 
the phase diagram for SC and SDW states, however 
this assumption requires further verifications. 

Using the forms of SC and SDW order parameters, 
we write the free and interaction parts in quadratic forms as 
\begin{widetext}
\be
\cH_0 =\onehalf \sum_{\vk}\left[ 
\xi_c(\vk)  \; c^\dag_{\vk\alpha} c_{\vk\alpha} + 
\xi_c(-\vk)  \; c^\dag_{-\vk\alpha} c_{-\vk\alpha} + 
\xi_f(\vk+\vq)  \; f^\dag_{\vk+\vq \alpha} f_{\vk+\vq \alpha} + 
\xi_f(-\vk-\vq)  \; f^\dag_{-\vk-\vq \alpha} f_{-\vk-\vq \alpha} 
\right] \,,
\ee
\be
\cH_\Delta = \onehalf \sum_{\vk} \left[ 
\Delta_c(\vk)_{\alpha\beta} \;  c^\dag_{\vk\alpha} c^\dag_{-\vk\beta} + 
\Delta_c^{\dag}(\vk)_{\alpha\beta}  \; c_{-\vk\alpha} c_{\vk\beta}  + 
\Delta_f(\vk+\vq)_{\alpha\beta}  \; f^\dag_{\vk+\vq\alpha} f^\dag_{-\vk-\vq\beta} +
\Delta_f^\dag(\vk+\vq)_{\alpha\beta}  \; f_{-\vk-\vq \alpha} f_{\vk+\vq \beta} 
\right] \,,
\ee
\be
\cH_m = \onehalf \sum_{\vk} \; \left[ 
m_{\vq, \alpha\beta} \; f^\dag_{\vk+\vq \alpha} c_{\vk\beta} + 
m_{\vq, \alpha\beta} \; c^\dag_{-\vk \alpha} f_{-\vk-\vq \beta} + 
m^\dag_{\vq, \alpha\beta} \; c^\dag_{\vk \alpha} f_{\vk+\vq \beta} +
m^\dag_{\vq, \alpha\beta} \; f^\dag_{-\vk-\vq \alpha} c_{-\vk \beta} 
\right] \,.
\ee
The Hamiltonian \req{eq:cHfull} can be represented in the matrix form
\be
\cH = \onehalf \sum_{\vk\alpha\beta} \;
\overline{\Psi}_{\vk\alpha} \, \widehat\cH_\vk \, \Psi_{\vk\beta} \,, 
\qquad
\widehat\cH_\vk =
\left( \begin{array}{c|c}
\begin{array}{cc} \xi_c(\vk) &   \Delta_c (\vk) \; i \sigma^y_{\alpha\beta} \\
-\Delta_c^* (\vk) \; i\sigma^y_{\alpha\beta} & -\xi_c(-\vk) \end{array}
&
\begin{array}{cc} m_\vq^* \; (\hat\vm \vsigma)^\dag_{\alpha\beta} & 0 \\
0 & - m_\vq^* \;  (\hat\vm \vsigma^\sm{T})^\dag_{\alpha\beta} \end{array}
\\ \hline
\begin{array}{cc} m_\vq \; (\hat\vm \vsigma)_{\alpha\beta} & 0 \\
0 & - m_\vq \; (\hat\vm \vsigma^\sm{T})_{\alpha\beta} \end{array}
&
\begin{array}{cc} \xi_f(\vk+\vq) &  \Delta_f (\vk+\vq) \; i\sigma^y_{\alpha\beta} \\
-\Delta_f^* (\vk+\vq) \; i\sigma^y_{\alpha\beta} & -\xi_f(-\vk-\vq) \end{array}
\end{array} \right)
\,,
\label{eq:Hk}
\ee
\end{widetext}
with 
$\overline{\Psi}_{\vk\alpha} =
(c^\dag_{\vk\alpha}, c_{-\vk\alpha}, f^\dag_{\vk+\vq\alpha}, f_{-\vk-\vq\alpha})$,
and $\Psi$ being its conjugated column. 
The two diagonal blocks of the matrix $\hat\cH_\vk$
correspond to a purely SC system with $\Delta_c$ 
and $\Delta_f$ living on two different bands, 
and two off-diagonal blocks contain SDW field $m_\vq$ 
that couples fermions between the two bands. 

To solve this system of equations for the SC and SDW order parameters,
Eqs.~(\ref{eqs:OP}), we define the imaginary-time Green's function 
\be
\whG(\vk,\tau)_{\alpha\beta} =
-\langle T_\tau \Psi(\tau)_{\vk\alpha} \overline\Psi(0)_{\vk\beta}
\rangle 
\equiv 
\left(\begin{array}{cc} \hat{G}_{cc} & \hat{G}_{cf} \\ 
\hat{G}_{fc} & \hat{G}_{ff} \end{array}\right)
\,,
\ee
which satisfies the Dyson equation, 
\be 
\whG^{-1}(\vk,\vare_n) = i\vare_n - \widehat\cH_\vk \,,
\label{eq:Dyson}
\ee
where $\vare_n=\pi T(2n+1)$ are the  Matsubara frequencies. 
The system of equations is closed by the self-consistency 
equations for the SC and SDW order parameters 
in terms of this Green's function, 
\be
\Delta_c(\vk) = \sum_{\vp} V^{sc}_{\vk,\vp}  \; 
T\sum_{\vare_n} \Tr{ (i\sigma^y)^\dag \hat \tau_+ \hat G_{ff}(\vp,\vare_n) } 
\label{eq:scSC}
\ee
\be
\Delta_f(\vk) = \sum_{\vp} V^{sc}_{\vk,\vp}  \; 
T\sum_{\vare_n} \Tr{ (i\sigma^y)^\dag \hat \tau_+ \hat G_{cc}(\vp,\vare_n) } 
\ee
\be
m_\vq  = - \sum_{\vp} \; V^{sdw} \frac{T}{4} \sum_{\vare_n}\; 
\Tr{ (\hat\vm \vsigma_4) \tauz \hat{G}_{fc}(\vp, i\vare_n )}.
\label{eq:scSDW}
\ee
Henceforth we define Pauli matrices in particle-hole space, 
$\hat\tau_{1,2,3}$, $\hat\tau_\pm=(\hat \tau_1\pm i\hat\tau_2)/2$, 
and the following matrices in spin- and particle-hole space, 
\be
\hDelta = \left(\begin{array}{cc} 0 & (\Delta \; i\sigma^y)_{\alpha\beta} \\ 
(\Delta \; i\sigma^y)^\dag_{\alpha\beta} & 0 
\end{array}\right) 
\quad, \qquad
\vsigma_4 = \left(\begin{array}{cc} \vsigma & 0 \\ 0 & \vsigma^\sm{T} 
\end{array}\right) \,,
\ee

The expressions above are valid for 
complex $\Delta(\vk)$ and $m_\vq$.
Below, to simplify 
formulas, we assume 
that $\Delta$'s and $m_\vq$
are real, i.e., consider only 
``sinusoidal'', $\cos\vQ\vR$, variations of the SDW order parameter. 
To lighten the notations, we will also drop the momenta
arguments $(\vk,\vk+\vq)$ in $\xi_{c,f}$, $\Delta_{c,f}$ and the subscript in
$m_\vq$ [still implying this dependence as it appears in \req{eq:Hk}]. 

The equations for components of the Green's function are obtained from 
inversion of \req{eq:Dyson}, 
\begin{subequations}
\bea
& \hat{G}_{cc}^{-1} = \hat{G}_{c0}^{-1} - m^2 \hat{G}_{f0} \,, 
\qquad
\hat{G}_{fc} = \hat{M} \hat{G}_{f0} \hat{G}_{cc} \,,
\\
& \hat{G}_{ff}^{-1} = \hat{G}_{f0}^{-1} - m^2 \hat{G}_{c0} \,,
\eea
\label{eq:Ginv}
\end{subequations}
with definition
\be
\left(\begin{array}{cc} 
\hat{G}_{c0}^{-1} & -\hat{M} \\ -\hat{M} & \hat{G}_{f0}^{-1} \end{array}\right) 
\equiv
\left( \begin{array}{c|c}
i\vare_n - \xi_c \tauz - \hDelta_c & -(\vm \vsigma_4)\tauz 
\\ \hline 
-(\vm\vsigma_4)\tauz & i\vare_n - \xi_f\tauz - \hDelta_f 
\end{array} \right) \,.
\label{eq:GMmatrix}
\ee
To obtain \req{eq:Ginv} we used the fact that the magnetic matrix $\hat{M}$ commutes with 
purely superconducting parts,
$[\hat{M}, \hat{G}_{c0}]=[\hat{M}, \hat{G}_{f0}]=0$, and $\hat{M} \hat{M} = m^2$. 

The diagonal Green's functions $\hat{G}_{c0}$ and $\hat{G}_{f0}$ 
are the same as in  a pure superconductor, e.g.,
\be
\hat{G}_{f0}(\vare_n) = \frac{\hat{G}_{f0}^{-1}(-\vare_n)}{D_{f0}} 
\,, \qquad
D_{f0} = \vare_n^2 + \xi_f^2 + \Delta_f^2  \,,
\ee
where for inversion we used the relations 
\be
\{\tauz, \hDelta\} = 0 \,,\quad 
[ (\hat\vm\vsigma_4) \tauz, \hDelta] = 0 \,, \quad 
\hDelta^2 = \Delta^2 \,,
\ee 
which are also employed to invert $4\times 4$ matrices for mixed SC+SDW state.  
For example, for $\hat{G}_{cc}$ we have 
\be
\hat{G}_{cc}(\vare_n) 
= \frac{1}{\hat{G}_{c0}^{-1}(\vare_n) - m^2 \hat{G}_{f0}^{-1}(-\vare_n)/D_{f0}} \,,
\label{n_1}
\ee 
and with the above relations in mind it becomes 
\begin{subequations}
\be
\hat{G}_{cc} = \hat{G}_{cc}^{(1)} + \hat{G}_{cc}^{(\tau_3)} + \hat{G}_{cc}^{(\Delta)} \,,
\ee
where
\bea
\hat{G}_{cc}^{(1)} = \frac{-i\vare_n( D_{f0} + m^2)}{D} \,,
\\
\hat{G}_{cc}^{(\tau_3)} = - \frac{\xi_c D_{f0} -\xi_f m^2}{D} \tauz \,,
\\
\hat{G}_{cc}^{(\Delta)} = - \frac{\hDelta_c D_{f0} -\hDelta_f m^2}{D} \,,
\eea
\label{eq:Gcc}
\end{subequations}
The denominator 
\bea 
&&\lefteqn{D = } 
\nonumber \\ 
&& \frac{\vare_n^2(D_{f0} + m^2)^2 + (\xi_c D_{f0} -\xi_f m^2)^2 
+(\Delta_c D_{f0} -\Delta_f m^2)^2}{D_{f0}} 
\nonumber \\ 
&&= (\vare_n^2 + \xi_c^2 + \Delta_c^2)(\vare_n^2 + \xi_f^2 + \Delta_f^2) 
\nonumber \\ 
&& \hspace{3cm} +2 m^2 (\vare_n^2 - \xi_c \xi_f -\Delta_c \Delta_f) + m^4 
\nonumber \\
&&= (\vare_n^2 + E_+^2) (\vare_n^2 + E_-^2)
\label{nn_1}
\eea
gives the energies of new excitations in the system, c.f. Ref.~\onlinecite{Fernandes09}.
We obtained (explicitly showing $\vk$ and $\vq$ here): 
\bea
\label{eq:excit}
E_\pm^2 &=& \xi_{\vk \vq}^2 + \delta_{\vk \vq}^2 + m^2 
	+ (\Delta_{\vk \vq}^-)^2 + (\Delta_{\vk \vq}^+)^2 
\\ \nonumber
&\pm &  2 \sqrt{m^2 [(\Delta_{\vk \vq}^+)^2 + \delta_{\vk \vq}^2] 
+ (\Delta_{\vk \vq}^-\Delta_{\vk \vq}^+ + \xi_{\vk \vq}\delta_{\vk \vq})^2 } \,,
\eea
with 
\bea
&& \xi_{\vk \vq} = \frac{\xi_f(\vk+\vq) - \xi_c(\vk)}{2} \,, \\
&&  \delta_{\vk \vq} = \frac{\xi_f(\vk+\vq) + \xi_c(\vk)}{2} \,, 
\eea
and 
\bea   
&& \Delta_{\vk \vq}^- = \frac{\Delta_f(\vk+\vq) - \Delta_c(\vk)}{2} \,
 \\
 &&  \Delta_{\vk \vq}^+ = \frac{\Delta_f(\vk+\vq) + \Delta_c(\vk)}{2}.
\label{nnnn_1}
\eea
The parameter $\xi_{\vk \vq}$ describes the dispersion and 
parameter $\delta_{\vk \vq}$ describes deviations of the electron and hole FSs 
from perfect nesting, as illustrated in 
Figs.~\ref{fig:FS} and \ref{fig:excitations}. 

For the inter-band part of the Green's function we obtain, 
\bea
\hat{G}_{fc}(\vare_n) 
&=& \hat{M} \hat{G}_{f0} \hat{G}_{cc} = 
\nonumber \\
&=& \hat{M} \hat{G}_{f0}(\vare_n) 
\frac{\hat{G}_{c0}^{-1}(-\vare_n) D_{f0} - m^2 \hat{G}_{f0}^{-1}(\vare_n)}{D} 
\nonumber \\
&=& \hat{M} \frac{\hat{G}_{f0}^{-1}(-\vare_n) \hat{G}_{c0}^{-1}(-\vare_n) - m^2}{D} \,,
\eea
where for self-consistency \req{eq:scSDW} we need only the part proportional 
purely to $\hat{M}$-matrix, \req{eq:GMmatrix},
\be
\hat{G}_{fc}^{({M})} = (\vm \vsigma_4)\tauz 
\frac{-\vare_n^2 + \xi_f \xi_c + \Delta_f \Delta_c - m^2}{D} \,.
\label{eq:Gfc}
\ee
Expressions for $\hat{G}_{ff}$ and $\hat{G}_{cf}$ are obtained from 
\req{eq:Gcc} and \req{eq:Gfc} 
by swapping indices, $c \leftrightarrow f$. 

We substitute the above expressions for
 $\hat{G}_{cc}^{(\Delta)}(\vp)$, 
$\hat{G}_{ff}^{(\Delta)}(\vp+\vq)$  and $\hat{G}_{fc}^{({M})}(\vp)$ 
 into
 the self-consistency equations \reqs{eq:scSC}-(\ref{eq:scSDW}) and arrive at
\begin{widetext}
\be
\Delta_f(\vk) = T\sum_{\vare_n} \sum_\vp (-2 V^{sc}_{\vk\; \vp}) 
\frac{\Delta_c(\vp) [\vare_n^2 + \xi_f^2(\vp+\vq) + \Delta_f^2(\vp+\vq)] 
- \Delta_f(\vp+\vq) \; m^2}{D} \,,
\label{eq:scSCgf}
\ee
\be
\Delta_c(\vk) = T\sum_{\vare_n} \sum_\vp (-2 V^{sc}_{\vk\; \vp+\vq}) 
\frac{\Delta_f(\vp+\vq) [\vare_n^2 + \xi_c^2(\vp) + \Delta_c^2(\vp)] 
- \Delta_c(\vp) \; m^2}{D} \,,
\label{eq:scSCgc}
\ee
\be
m = T\sum_{\vare_n} \sum_\vp V^{sdw} 
 \frac{\vare_n^2 + m^2 -  \xi_f(\vp+\vq)\xi_c(\vp) - \Delta_f(\vp+\vq) \Delta_c(\vp)}{D} m \,.
\label{eq:scSDWgf}
\ee
\end{widetext}

To calculate the relative stability of different states, 
one also needs to evaluate the free energy. We follow the 
Luttinger-Ward\cite{lutt-ward} and De Dominicis-Martin\cite{dominicis-martin} 
method, and consider the functional\cite{rainer76}
\be
F = -\onehalf Sp \left\{ 
	\whSigma \whG + \ln [ -(i\vare_n - \hat\xi) + \whSigma ] \right\} 
	+ \Phi[\whG] \,,
\label{eq:luttward}
\ee
which, if minimized with respect to $\whG$, gives self-consistency 
equations, $\whSigma[\whG] = 2\delta\Phi[\whG]/\delta\whG$; 
and, if minimized  with respect to $\whSigma$, gives the Dyson equation, \req{eq:Dyson}. 
Here $Sp$ is 
the trace over two fermion bands, spin, 
particle-hole matrix structure, and the 
sum over Matsubara energies and the integral over momenta, 
and $\whSigma$ is the mean field SC and SDW order parameter matrix, 
\be
\whSigma = \left(\begin{array}{cc} \hDelta_c & \hat{M} \\ 
\hat{M} & \hDelta_f  
\end{array}\right)  \,.
\label{eq:whSigma}
\ee
The functional $\Phi[\whG]$ producing the self-consistency 
equations is a quadratic function of $\whG$.
Using the self-consistency equations 
one can explicitly verify that 
at weak-coupling it can be written as 
$\Phi[\whG] = \onefourth Sp \{ \whSigma \whG\} $.
To deal with the logarithm in \req{eq:luttward} one introduces 
a continuous variable $\lambda$ instead of $\vare_n$, differentiates the logarithmic term
with respect to $\lambda$ to obtain the Green's function 
$ \whG(\lambda)=\left( i\lambda-\hat \xi-\hat\Sigma \right)^{-1} $, 
and then
integrates back to get the difference between 
a condensed state and the normal state for fixed external 
parameters, such as temperature or field,   
\be
\Del F (\Delta_{c,f},m) 
= -\onehalf Sp 
\left\{ \onehalf \whSigma \whG 
- \int\limits_{\vare_n}^{\infty} d\lambda [ i\whG(\lambda) - i\whG_N(\lambda)] 
\right\}
\label{eq:FEgeneral}
\ee
where $\whG_N$ is the Green's function in the normal state 
without either SC or SDW order,
and we used the fact that in the normal state $\whSigma =0$.
Substituting 
into (\ref{eq:FEgeneral}) the Green's functions \reqs{eq:Gcc}, (\ref{eq:Gfc}), 
the self-energy \req{eq:whSigma}, and using the self-consistency equations 
\reqs{eq:scSCgf}-(\ref{eq:scSDWgf}) to 
eliminate the high-energy cut-offs in order to regularize 
the $\vare_n$-summation and $\vk$-integration, 
one obtains the most general free energy functional 
for given $\Delta_{c,f}$ and $m$. 

\subsection{\label{sec:smallFSsplit} Limit of small Fermi surface splitting}

In principle, equations for full Green's functions 
\req{eq:Gcc}, (\ref{eq:Gfc}), the self-consistency equations 
\reqs{eq:scSCgf}-(\ref{eq:scSDWgf}) and the free energy \req{eq:FEgeneral}, 
completely describe the system in a very general case.
 However, to proceed further with the analytics one can 
reduce the number of summations which is also desirable 
from a numerical standpoint. 

The typical approximation is to 
linearize the dispersion near the FS  and 
integrate out the momenta 
in the direction normal to the FS over $\xi_{\vk \vq}$.
In the case, when the two FSs are reasonably close 
to each other (when shifted by $(0,\pi)$),  
and electron and hole dispersions are similar, 
the values of FS mismatch $\delta_{\vk \vq}$ 
are weakly momentum dependent, and can be taken at 
positions where $\xi_{\vk \vq} = 0$.

The consequence of this approximation,
which we discuss in some detail in Appendix \ref{appA},
is that $\delta_{\vk \vq}$ 
depends only on the angle in $\vk$-space, but not on $\xi_{\vk \vq}$ 
and hence one can integrate along 
a particular direction $\hvk$ over $\xi_{\vk \vq}$, 
keeping $\delta_{\hvk \vq}$ fixed. 
 
Within this approximation the DOS for both FSs are
the same, and the magnitudes of $\Delta_c$ and $\Delta_f$ are equal
(the angular dependence of SC gaps is still 
determined by that of the SC interactions).
There are, indeed, also higher order terms, which we neglected in the last lines
of \req{nn_2}. These terms  make hole and electron DOS different from each
other, what in turn makes $|\Delta_c|$ and $|\Delta_f|$ non-equal, but these
terms are small in $\delta_\vk/\mu_{c,f}$
and only account for sub-leading terms in
the free energy, $\mu_{c,f}$ are Fermi energies of electron and hole bands, \req{eq:xis}.

This approximation comes at certain price. 
When two FSs are of very different 
shapes, approximating them as small deviations from a single 
line in $\vk$-space everywhere is incorrect. 
This is shown for example in \rfig{fig:excitations}(d), where
the two FSs are quite different away from the crossing points. 
 However in this case one realizes that if at some 
$\vk$-point the two bands are far apart, 
the effect of the SDW is very small, and we can approximate 
those FS parts as participating in SC pairing only, 
with little or no competition from the SDW interaction.  
This can be seen from \req{eq:Ginv} for the Green's function. 
For example, for electrons near the FS of the $c$-band, $\xi_c \to 0$, 
$\xi_f$ is large and $\hat{G}_{cc}^{-1} \approx \hat{G}_{c0}^{-1} + \cO(m^2/\xi_f)$, 
and the corrections due to $m$ can be neglected 
when we go along $c$-FS 
away from the region where $\xi_c \approx \xi_f \approx 0$. 
We will return to this issue in section \ref{sec:Nsdw}, 
to show that the results are qualitatively the same 
whether we consider large or small splitting of the FSs. 

For small splitting between hole and electron Fermi surfaces, 
we perform $\xi$-integration analytically. For this 
we approximate  $V^{sc}_{\vk,\vp}$ by an isotropic  $V^{sc}$, i.e., 
take angle-independent SC gap. The sign of $V^{sc}$ can be arbitrary,
and we consider separately the two cases: 

a) $V^{sc} >0$: 
results in the 
 $s^\es$ state, with gaps of opposite signs for electrons and holes, 
$$\Delta_f = -\Delta_c = \Delta  \, \quad \mbox{or} \quad \Delta_+=0, \quad \Delta_-=\Delta ;$$

b) $V^{sc} <0$: $s^\pp$ state, with the same gaps on two FSs, 
$$\Delta_f = \Delta_c = \Delta \,\quad \mbox{or} \quad \Delta_+=\Delta, \quad \Delta_-=0  \,. $$

In both cases $\Delta_+ \Delta_-= 0$ and the denominator of the Green's 
function can be written as, 
\be
D = (\vare_n^2+E_+^2)(\vare_n^2+E_-^2) 
  = (\xi_{\vk \vq}^2+\Sigma_+^2)(\xi_{\vk \vq}^2+\Sigma_-^2) 
\,, 
\ee
where 
\be
\Sigma_{\pm}^2 = \vare_n^2 + \Delta^2 + m^2 -\delta^2_{\hvk \vq}
	\pm 2\sqrt{ m^2 \Delta^2 \frac{1+s}{2} -\delta^2_{\hvk \vq} (\vare_n^2 + \Delta^2) } 
\label{ch_1}
\ee
with $s=+1$ ($s=-1$) corresponding to $s^{\pp}$ ($s^\es$) state.
Closing the integration contours over $\xi_{\vk \vq}$ 
in the self-consistency equations and in the free energy
over the upper half-plane
and counting poles at $+i\Sigma_\pm$ we obtain
\begin{widetext}
\bea
&& \frac{-s}{v^{sc}} \Delta 
= \pi  T \sum_{|\vare_n|<\Lambda} 
\left\langle \frac{\Delta}{\Sigma_++\Sigma_-}
\left(1 + \frac{\vare_n^2 + \Delta^2 +\delta_{\hvk \vq}^2 - s \, m^2}
{\Sigma_+ \Sigma_-} \right) 
\right\rangle \,,
\qquad \frac{1}{|v^{sc}|} = \ln \frac{1.13 \Lambda}{T_c} \,,
\label{eq:scSCqc}
\\
&& \frac{1}{v^{sdw}} m 
=  \pi T \sum_{|\vare_n|<\Lambda} 
\left\langle \frac{m}{\Sigma_++\Sigma_-}
\left(1 + \frac{\vare_n^2  + m^2 - \delta_{\hvk \vq}^2 - s \, \Delta^2}
{\Sigma_+ \Sigma_-} \right) 
\right\rangle 
\,,
\qquad \frac{1}{v^{sdw}} = \ln \frac{1.13 \Lambda}{T_s} \,,
\label{eq:scSDWc}
\\ 
&& \frac{\Del F(\Delta, m)}{4N_F} 
= \frac{\Delta^2}{2} \ln\frac{T}{T_c} + \frac{ m^2}{2} \ln\frac{T}{T_s} 
-  2\pi T \sum_{\vare_n>0} 
\left\langle \onehalf(\Sigma_+ + \Sigma_-)
- \vare_n - \frac{\Delta^2}{2\vare_n} - \frac{m^2}{2\vare_n}
\right\rangle \,,
\label{eq:feqc}
\eea
\end{widetext}
where angle brackets denote remaining momentum averaging over directions on the FS.
$N_F$ is the density of states at the FS per spin,
and $v^{sc}=2N_F V^{sc}$ and $v^{sdw} = N_F V^{sdw}$ 
are the dimensionless couplings in the
SC and SDW channels.\cite{Chubukov08}
Taken alone, $v^{sc}$ leads to a SC state with critical temperature $T_c$,
which is independent of $\delta_{\hvk \vq}$ as one can 
 check by setting $m=0$ in \req{eq:scSCqc}, 
while $v^{sdw}$ leads to an SDW state with transition temperature $T_s$ 
which does depend on $\delta_{\hvk \vq}$. 
We define $T_s$ as the SDW transition temperature 
at \emph{perfect} nesting, when $\delta_{\hvk \vq}\equiv 0$. 

The relative sign between SC and SDW orders, as given by 
terms $-s \, \Delta \, m^2$ in \req{eq:scSCqc} 
and $-s \, m \, \Delta^2$ in \req{eq:scSDWc}, 
is positive for $s^\es$ state resulting in effective ``attraction''
of the two orders, and negative for $s^\pp$ state
implying that the formation of one order resists the  
appearance of the other.\cite{Vavilov09gl,Fernandes09}
The actual co-existence of 
the two orders, however, 
is a more subtle effect and 
needs to be determined from 
the exact solution of these equations and the analysis of
the free energy.
 The difference in excitations energies \req{eq:excit} and \req{ch_1} 
between $s^\pp$ and $s^\es$ states is also consistent with previous studies of 
$d$- and $p$- wave superconductivity in heavy-fermion 
metals,\cite{MachidaKato87} 
that concluded the SC states with symmetries 
$P \, T_Q = -1$ (e.g. $s^\es$, $d$), where  
$P$ is parity [$P \Delta(\vp)=\Delta(-\vp)$] and 
$T_Q$ is the shift by the nesting vector [$T_Q \Delta(\vp)=\Delta(\vp+\vQ)$],  
are more likely to form co-existence with SDW than 
those with $P \, T_Q = +1$ (e.g. $s^\pp$, $p$). 

We will also analyze the quasiparticle density of states (DOS), 
which is given by the integrals over $\xi_{\vk \vq}$ 
of the diagonal components of the Green's function.
For example 
for $c$-fermions
\bea
g_c(\vare_n, \hvk) &=& \int \frac{d\xi_{\vk \vq}}{\pi} G^{(1)}_{cc}  
\\ \nonumber
&=& \frac{ -i\vare_n}{\Sigma_+ + \Sigma_-} \left(
1+ \frac{\vare_n^2 + m^2 + \Delta^2 + \delta_{\hvk \vq}^2}{\Sigma_+ \Sigma_-}
\right) \,,
\eea
which for pure SDW state reduces to 
\be
g_c(\vare_n, \hvk) = -\onehalf 
\sum_\pm 
\frac{ i\vare_n \pm  \delta_{\hvk \vq}}
{\sqrt{m^2 - (i\vare_n \pm \delta_{\hvk \vq})^2}}  \,,
\label{eq:dosSDW}
\ee
and actual DOS is obtained by analytic continuation, 
\be
\frac{N(\epsilon, \hvk)}{N_F} = - \Im g(i\vare_n \to \epsilon + i 0^+, \hvk) \,.
\ee

\section{Pure SDW state}
\label{sec:pureSDW}

In pnictides, parent materials 
usually have only magnetic order 
below a transition temperature $T_s$. 
Superconductivity appears at a finite doping, 
when the SDW transition is suppressed. 
Keeping this in mind,  
we consider first a purely SDW state,
and analyze how it is modified
when FSs are deformed by addition 
or removal of electronic carriers, and 
whether modified FSs are still present in the SDW phase.
To remind, we denote by $T_s$ the SDW-N transition temperature 
at perfect nesting, which effectively gives the scale of SDW 
interaction in the system. The true instability temperature,
which we denote explicitly by $T_s(\delta_{\hvk \vq})$, 
is a function of ellipticity, doping, and incommensurability.

We begin by presenting explicit formulas for the excitation spectrum, the 
SDW order parameter, and the free energy.
For $\Delta =0$, 
$\Sigma^2_{\pm}$ given by (\ref{ch_1}) is
\be
\Sigma_{\pm}^2 = \left(\vare_n \pm  i\delta_{\hvk \vq}\right)^2 + m^2
\label{ch_2}
\ee
and the excitation spectrum consists of four branches with energies
$\pm E_\pm (\Delta=0)$, where  
\be
E_\pm(\Delta=0) = \sqrt{\xi_{\vk \vq}^2 + m^2} \pm \delta_{\hvk \vq} \,,
\label{eq:SDWexc}
\ee
In (\ref{ch_2}) and (\ref{eq:SDWexc})
\bea
\xi_{\vk \vq}&=& \frac{\xi_f(\vk+\vq) -\xi_c(\vk)}{2} \,, 
\nonumber \\
 \delta_{\hvk \vq} &=& \frac{\xi_f(\vk+\vq) + \xi_c(\vk)}{2} 
\approx  
\frac{\xi_f(\vk^c_F+\vq)}{2} \nonumber \\
 &=& \frac{v_F}{2} \hvk (\vk^c_F-\vk^f_F+\vq).
\label{ch_2_1}
\eea
We remind that $\delta_{\hvk \vq}$  describes the mismatch
between the shapes of the electron and hole bands and determines their nesting
properties in $\hvk$-direction. 

Equation~(\ref{eq:scSDWc}) for the SDW order parameter $m$ simplifies to
\beq
\frac{1}{v^{sdw}} = 2 \pi T \sum_{0<\vare_n<\Lambda} 
\Re \frac{1}{\sqrt{(\vare_n + i\delta_{\hvk \vq})^2 + m^2}},\\
\label{ch_3}
\ee
and the cut-off $\Lambda$ can be eliminated in favor of $T_s$, 
 \be
\ln \frac{T}{T_s} = 2 \pi T \sum_{\vare_n > 0} \;
\Re \left\langle \frac{1}{ \sqrt{(\vare_n+i\delta_{\hvk \vq})^2+m^2} }
-\frac{1}{|\vare_n|} \right\rangle \,,
\label{eq:scSDWpure}
\ee
where the summation over $\vare_n$ now extends to infinity. 
Second-order transition temperature $T= T_s(\delta_{\hvk \vq})$ 
is obtained by setting $m =0$:
\be
\ln \frac{T}{T_s} = 2 \pi T \sum_{\vare_n > 0} 
\Re \left\langle \frac{1}{\vare_n+i\delta_{\hvk \vq}} 
-\frac{1}{\vare_n} \right\rangle . 
\label{eq:ts} 
\ee
The free energy, Eq. (\ref{eq:feqc}),  becomes 
\bea
&& \frac{\Del F(m)}{4N_F}  = \frac{ m^2}{2} \ln\frac{T}{T_s}
\nonumber \\
&& -  2\pi T \sum_{\vare_n>0}\left( \Re \left\langle 
\sqrt{(\vare_n + i\delta_{\hvk \vq})^2 + m^2}\right\rangle - \vare_n -
\frac{m^2}{2\vare_n}\right) \nonumber \\
&&= \frac{ m^2}{2} \ln\frac{1.13 \Lambda}{T_s} 
\nonumber \\
&& -  2\pi T \sum_{0< \vare_n < \Lambda}\left( \Re \left\langle 
\sqrt{(\vare_n + i\delta_{\hvk \vq})^2 + m^2}\right\rangle - \vare_n\right) \,.
\label{ch_4}
\eea
 
Below we consider several special cases 
for $\delta_{\hvk \vq}$ (see \rfig{fig:excitations}): 

\begin{figure}[t]
\centerline{\includegraphics[width=0.8\linewidth]{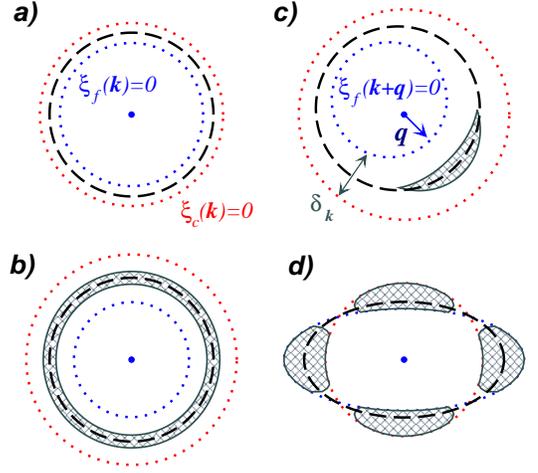}}
\caption{(Color online) 
	The appearance of gapless excitations in the presence of SDW order.  
	The dotted lines indicate FSs for electrons, $\xi_f=0$, and holes,
	$\xi_c=0$. The dashed curve is an ``effective'' FS, $\xi_{\vk \vq}=0$. 
	a) 
	when $q=0$ and $m$ is large compared to FS mismatch, $m > \delta_{\hvk 0}$, 
 	all excitations are gapped; 
	b) 
	when $m$ is small,
 	gapless excitations are preserved along the 
	two modified FSs at $\xi_{\vk,0} = \pm (\delta^2_{\hvk,0} -m^2)^{1/2}$ 
	($|\delta_{\hvk,0}| > m$ in the shaded region). Such gapless state, however, only
	exists at high temperatures, while at low $T$ it is 
	pre-emptied by a first order transition to the normal state~\cite{Vorontsov09magsc}; 
	c) to prevent the first order transition, magnetic order is formed at 
	an incommensurate vector $\vQ = \vQ_0 +\vq$. 
	This improves electron-hole nesting on some part of 
	the FS, but allows for gapless excitations at the opposite side; 
	d) when the two FSs are of different shapes, the nested parts 
	become gapped due to SDW order,
	and on the rest of the FSs the excitations are little 
	affected by SDW order. 
	The density of states for these  cases is shown in \rfig{fig:dosSDW}. 
}
\label{fig:excitations}
\end{figure}

$\bullet$ 
 two co-axial circles, $\vq=0$: $\vk^c_F-\vk^f_F = \hvk (k^c_F - k^f_F)$:
\be 
\delta_{\hvk \vq} =  \onehalf v_F |k^c_F - k^f_F| \equiv \delta_0. 
\label{ch_n1}
\ee
For a fixed $\delta_0$, 
circular hole and electron FSs survive in the SDW phase when $m < \delta_0$ 
(\rfig{fig:excitations}b),
but come closer to each other as $m$ increases and merge at $m = \delta_0$.
At larger $m$ all excitations are gapped  (\rfig{fig:excitations}a).  

$\bullet$ 
FS of different shapes, e.g., one circle and one ellipse, co-centered:  
$\vq=0$ with 
$\vk^c_F-\vk^f_F = \hvk (k^c_F - k^f_F + \Del k \, \cos 2\phi)$:
\be 
\delta_{\hvk \vq} = \delta_0 + \delta_2 \cos2\phi \,, \qquad 
\delta_2 = \onehalf v_F \Del k
\label{ch_n2}
\ee
In this case, at small enough $m$, the FS has a form of two hole and
two electron pockets. As $m$ gets larger, the pockets shrink and eventually disappear.

\begin{figure}[t]
\centerline{\includegraphics[width=0.8\linewidth]{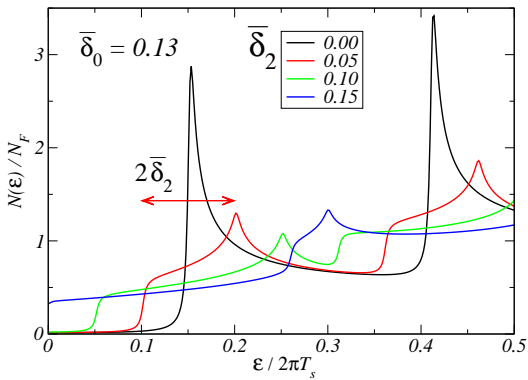}}
\caption{(Color online) 
	The FS averaged DOS for pure SDW state and FS mismatch 
	$\delta_{\hvk \vq} = \delta_0 + \delta_2 \cos 2\phi$. 
	We use dimensionless variables denoted by bars, 
	$\bar \delta_0 = \delta_0/2\pi T_s = 0.13$, 
	zero-temperature SDW gap $\bar{m}_0 = m(T=0)/2\pi T_s = 0.28$ 
	and vary $\bar \delta_2=\delta_2/2\pi T_s$. 
	For $\bar \delta_2 =0$, $N(\epsilon)$ vanishes below 
	$\bar \epsilon = \bar{m}-\bar\delta_0$
	and has two sharp BCS peaks at $\bar \epsilon = \bar{m} \pm \bar\delta_0$. 
	At finite $\bar\delta_{2}$, each of the two peaks splits into a ``band'' 
	bounded by two weaker non-analyticities separated by $2\bar\delta_2$.  
	The gap in the DOS behaves as $\bar{m} - \bar\delta_0 - \bar\delta_2$ 
	and closes when $\bar\delta_0 + \bar\delta_2 \ge \bar{m}$, 
	metallic states forms. 
	The DOS remains the same if we replace the ellipticity parameter 
	$\bar\delta_2$ by the incommensurability parameter $\bar{q}$.
	}
	\label{fig:dosSDW} 
\end{figure}

$\bullet$
 two circles
of different radii, centers shifted by $\vq$: 
\be
 \delta_{\hvk \vq} = \delta_{\hvk} = 
\delta_0 + \onehalf \vv_F \vq 
= \delta_0 + \onehalf v_F q  \cos(\phi-\phi_0)
\label{ch_n3}
\ee
 where $\phi$ and $\phi_{0}$ are the directions of $\vv_F$ and $\hat\vq$. 
In this case, when $m$ increases,  
gapless excitations survive along a pocket 
in one region of the $\vk$, while excitations with $-\vk$ become gapped
( \rfig{fig:excitations}c). 
At large enough $m$, modified FS disappears and excitations with all momenta
$\vk$ become gapped.
This scenario refers to the case when the magnetic 
ordering occurs at a vector, different from the nesting 
vector $\vQ_0$, producing incommensurate SDW state. 
It may occur because the electronic system has an option to choose 
$\vq\ne0$ if it minimizes the energy, 
or because the SDW interaction is peaked at a fixed $\vQ \ne \vQ_0$ 
for some reason. 
Note that \req{eq:scSDWpure} for SDW order is a magnetic analog 
of Fulde-Ferrell-Larkin-Ovchinnikov (FFLO) state\cite{FFLO} 
in a paramagnetically limited superconductor.
An incommensurate SDW state at finite dopings has been studied
in application to chromium and its alloys\cite{rice70,kulikov84,Fawcett94reviewSDW} 
and, more recently, to pnictide 
materials.~\cite{Cvetkovic09vdw,Vorontsov09magsc}

In general, all three terms are present, and 
\be
\delta_{\hvk \vq} =\delta_0+  \delta_2 \cos 2\phi + \onehalf v_F q 
\cos(\phi-\phi_0) \,. 
\ee
In the figures we use dimensionless 
parameters, that are denoted by a bar. 
For isotropic and anisotropic FS distortions, 
\be
{\bar\delta}_{0,2} = \frac{\delta_{0,2}}{2\pi T_s} \,, \qquad
{\bar q} = \frac{v_F q}{4\pi T_s}\,, 
\label{ch_15}
\ee
and similarly for other energy variables,
\be
\bar{m} = \frac{m}{2\pi T_s} \,, \quad 
\bar{\epsilon} = \frac{\epsilon}{2\pi T_s} \,, \quad 
\bar{\Delta} = \frac{\Delta}{2\pi T_c} \,.
\label{ch_16}
\ee
We use different notations for 
prefactors of $\cos(\phi-\phi_0)$ and $\cos 2 \phi$ terms to
emphasize that they have different origin: $\delta_2$ is an ``input''
parameter defined by the elliptic form of the electron FS due to the electronic
band structure, while $q$ is adjustable parameter that minimizes the
free energy of the system. If the minimum of the free energy corresponds to
$q =0$, SDW order is commensurate, 
otherwise SDW order is incommensurate.

In \rfig{fig:dosSDW} we show the DOS $N({\epsilon})$ for the fixed 
$\bar \delta_0 = 0.13$ and 
 $\bar{m}  = 0.28$, and different $\delta_2$. 
For $\delta_2=0$, $N(\epsilon)$ vanishes below $\epsilon = m-\delta_0$ 
and has two BCS-like peaks at $\epsilon = m \pm \delta_0$. 
At finite ${\delta}_{2}$, each of the two peaks 
spreads into a region of width $2\delta_2$ bounded by 
two weaker non-analyticities. The gap in the DOS behaves
as $m - \delta_0 - \delta_2$ and closes when $\delta_0 + \delta_2$ 
become larger than $m$. 
The DOS and all other results remain the same if we replace 
the ellipticity parameter $\bar\delta_2$ by the incommensurability 
parameter $\bar{q}$ because the angular integral in \req{eq:dosSDW} or \req{eq:ts} 
over momentum directions on the FS 
coincides for $\cos(\phi-\phi_0)$ and $\cos2\phi$ terms 
in $\delta_{\hvk,\vq}$, if considered separately. 
 The DOS and $T_s(\delta_{\hvk \vq})$ change, however, 
when both $\delta_2$ and ${q}$ are present simultaneously.

Below we discuss the phase diagram for the pure SDW state to the extend that we
will need to analyze  potential co-existence between SDW and SC states, which
is the subject of this paper. 

It is instructive to consider separately the case when SDW order is set to
remain commensurate for all $\delta_{0,2}$ (i.e., $q =0$), 
and the case when the system can choose $q$. 
In our model, the first case is 
artificial and just sets the stage to
study the actual situation when the value of 
$q$ is obtained by minimizing the free energy. 
However, a commensurate magnetic order may
 be stabilized in the SDW state,
if the interaction $V^{sdw}$ is by itself sharply peaked at the 
commensurate momentum $\vQ_0$.

\begin{figure}[t]
\centerline{\includegraphics[width=\linewidth]{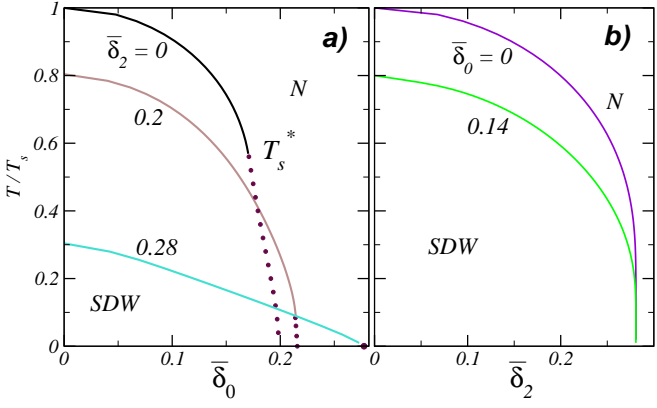}}
\caption{
	(Color online) The SDW-N transition for commensurate SDW order. 
	The parameters ${{\bar \delta}}_0$ and ${\bar \delta}_2$ describe
	the difference between the area of hole and electron pockets and the
	ellipticity of the electron pocket, respectively.  
	Here and in all subsequent figures dotted lines mark first-order transitions, 
	solid and dashed lines mark second-order transitions. 
		Panel (a): 
	variation of the transition temperature with $\delta_0$ 
	for fixed $\delta_2$. 
	The transition is second order at small $\delta_0$ 
	but becomes first order at larger $\delta_0$. 
	At $\delta_2 =0$, the transition becomes first order 
	at $T^*_s \approx 0.56 T_s$.  
		Panel (b): 
	variation of the transition temperature with $\delta_2$ for fixed 
	$\delta_0$. $T_s(\delta_0, \delta_2)$ monotonically decreases with
	increasing $\delta_2$ and vanishes at the same value 
	${\bar \delta}_2 \approx 0.28073$, independent of $\delta_0$.  
	} 
	\label{fig:sdw} 
\end{figure}

The results for the case $q \equiv 0$ are presented in  \rfig{fig:sdw}.
In panel (a) we present the results for the transition temperature 
$T_s (\delta_0,\delta_2)$ for several values of $\delta_2$. 
All curves show that the transition is second-order at high $T$ and first-order
at small $T$. The first-order transition lines 
(dotted lines in \rfig{fig:sdw}(a)) 
were obtained by solving numerically the nonlinear equation for $m$, 
substituting the result into the 
free energy (\ref{ch_4}) 
and  finding a location where
$\Del F(m) =0$. 

To verify 
that the transition becomes first order at low $T$, we expanded the 
 free energy in powers of $m$ as 
\be
\Del F(m) = \alpha_m m^2+B m^4 + \dots \,,
\label{eq:GLfemag} 
\ee
and checked the sign of the $B$ term. 
The coefficients $\alpha_m$ and $B$ are determined from \req{ch_4},  
\bea
&& \alpha_m = \onehalf \left(
\ln\frac{T}{T_s} - 2\pi T \sum_{\vare_n>0}
\Re \left\langle \frac{1}{\vare_n + i\delta_\hvk} -\frac{1}{\vare_n} \right\rangle \right)  \,,
\nonumber \\ 
&& B = \frac{\pi T}{4} \sum_{\vare_n>0}
\Re \left\langle \frac{1}{(\vare_n + i\delta_\hvk)^3}\right\rangle \,,
\label{eq:GLmag}
\eea
where $\delta_\hvk = \delta_0 + \delta_2 \cos2\phi$.
Solid lines in \rfig{fig:sdw}(a) correspond to $\alpha_m =0$. The N-SDW
transition is second order and occurs when $\alpha_m=0$ if $B >0$, but
becomes first order and occurs before $\alpha_m$ 
becomes negative if $B<0$. 
We indeed found that for 
{\it all} fixed $\delta_2$, 
for which  SDW-N transition is possible, 
$B$ changes sign along the line $\alpha_m =0$ and becomes negative at small $T$. 
For $\delta_2=0$, this occurs at $T^*_s = 0.56\, T_s$ and $\bar\delta_0^* = 0.17$.  

We point out the following counter-intuitive feature in \rfig{fig:sdw}(a). 
Increase in $\delta_2$ reduces the transition temperature 
at $\delta_0=0$, and at the same time makes the curve 
flatter allowing for a larger SDW region along $\delta_0$. 
The transition line becomes completely flat 
at a critical value 
$\delta_{2c} =  0.28073 (2\pi T_s)$ 
(see below) when $T_s (\delta_0, \delta_{2c})  = +0$. 
At this point, it spans the interval $\delta_0 \in [0, \delta_{2c}]$. 
The existence of the SDW ordered state at $\delta_2 = \delta_{2c}$ 
over a finite range of $\delta_0$ despite that the
transition temperature is $+0$ 
is a highly non-trivial effect which deserves a
separate discussion.~\cite{we_new}

In \rfig{fig:sdw}(b) we show the transition temperature at fixed $\delta_0$, 
as a function of the ellipticity parameter $\delta_2$. 
As expected, $T_s (\delta_2)$ monotonically decreases with increasing
ellipticity of the electron band. 
The SDW order exists up to ${\delta}_{2c}$,
at which $T_s (\delta_{2c}) =+0$. 
The value of ${\delta}_{2c}$ is independent of $\delta_0$ and 
can be obtained  
by taking the limit $T \to 0$ in (\ref{ch_3}) with $m=0$ and
re-writing this equation as 
\be
\frac{1}{v^{sdw}} 
= \Re \int\limits_0^\Lambda d\vare 
\left\langle
\frac{1}{\vare + i \delta_{\hvk} } \right\rangle 
= \Re \ln\frac{2\Lambda}{i\delta_0+\sqrt{\delta_2^2 - \delta_0^2}} \,.
\label{eq:zeroTgap}
\ee
The interaction can be eliminated in favor of zero-temperature gap $m_0$ 
at $\delta_0 = \delta_2 =0$
\be\label{eq:m0}
\frac{1}{v^{sdw}} 
=  \int\limits_0^\Lambda d\vare 
\frac{1}{\sqrt{\vare^2 + m_0^2}}
= \ln\frac{2\Lambda}{m_0},
\ee
where from Eq. (\ref{eq:scSDWpure}) we obtain, at $\delta_0 = \delta_2 =0$: 
\be
m_0 = \frac{2\pi T_s}{2e^{\gamma_E}} =  0.28073 \times (2\pi T_s) 
\ee
and  $\gamma_E \approx 0.57722$  is 
Euler's constant.  
At finite $\delta_0$ and $\delta_2$, the value of  $m$ at $T=0$ remains equal
to $m_0$ as long as $\delta_0 + \delta_2 < m_0$.
The combination of Eqs. (\ref{eq:zeroTgap}) and (\ref{eq:m0}) gives  
$\delta_{2c} = m_0 = 0.28073\times (2\pi T_s) $,
provided that $\delta_0 < \delta_{2c}$ 
(there exists another solution 
$\delta_2^2 = 2\delta_0 m_0 -m_0^2$ at  $\delta_2<\delta_0$ and 
$\delta_0 >m_0/2$, but it corresponds to an unstable state).
A similar result has been obtained in 
the studies of FFLO transition.~\cite{FFLO,abrikosovMetals}

The form of $T_s (\delta_2)$ near $\delta_{2c}$ can be obtained analytically
by re-writing the condition $\alpha_m=0$ in (\ref{eq:GLmag}) as 
\be
\ln\frac{T}{T_s} + 2\pi T \sum_{\vare_n>0}
\left\langle \frac{\delta_2^2 \cos^2{2 \phi}}{\vare_n (\vare^2_n + \delta^2_2
 \cos^2{2\phi})} \right\rangle = 0, 
\ee
integrating explicitly over $\phi$, re-expressing 
$1/\sqrt{\vare^2_n + \delta^2_2}$ as 
$(2/\pi) \int_0^\infty dx/(x^2 + \vare^2_n + \delta^2_2)$, and
performing the summation over $\vare_n$ before the integration over $x$. 
Carrying out this procedure, we obtain
\be
T_s (\delta_2) \approx 
\frac{\delta_{2c} }{|\ln\left(1-{\delta_2}/{\delta_{2c}}\right)|}
\label{eq:sdwTs}
\ee
We see that $T_s$ very rapidly increases at deviations from $\delta_{2c}$. For
$\delta_2 =  0.9974 \delta_{2c}$ (${\bar \delta_2} =0.28$),
 Eq. (\ref{eq:sdwTs}) yields $T_s (\delta_2) \approx 0.3 T_s$, in good agreement
 with \rfig{fig:sdw}(a).

Also, one can easily show that at $T=0$ fermionic  excitations in the SDW 
state are all gapped when $m_0 > \delta_0 + \delta_2$. When $\delta_0 +
\delta_2 >m_0$, the SDW state possess Fermi surfaces and gapless fermionic excitations.  

\begin{figure}[t]
\centerline{\includegraphics[width=0.75\linewidth]{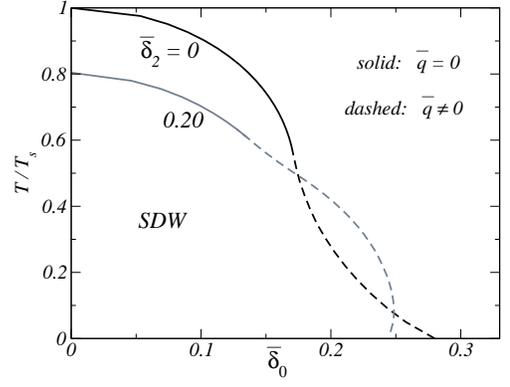}}
\caption{
	(Color online) Same as in \rfig{fig:sdw}, but when the system is allowed
	to choose between commensurate and incommensurate SDW orders. Solid lines
	are second-order transition lines into a state with a commensurate SDW
	order, dashed lines are second order transition lines into an SDW state
	with an incommensurate SDW order (the magnetic analog of FFLO state).
	For all $\delta_2 >0$, incommensuration occurs before the
	commensurate transition becomes first order 
	(the onsets of incommensuration and first-order transition coincide 
	for $\delta_2 =0$). 
	Observe that incommensuration develops at progressively smaller
	$\delta_0$ as $\delta_2$ increases and $T_s(\delta_0)$ decreases, 
	but the range of $\delta_0$ over which incommensurate SDW order exists
	actually increases with increasing $\delta_2$. 
	} 
	\label{fig:sdwFFLO} 
\end{figure}

We next consider the case when the system is free to choose between
commensurate and incommensurate SDW orders and may develop incommensurate order
to lower the free energy. In  \rfig{fig:sdwFFLO} we show the transition
temperature $T_s ({\delta}_0)$ for fixed ${\delta}_2$.  We found that,
for all ${\delta}_2$, first order transition is overshadowed by a
transition into an incommensurate SDW state. For ${\delta}_2=0$,
incommensuration develops exactly where $B$ changes sign, and the transition
into incommensurate SDW state remains second order for all ${\delta}_0$.
For ${\delta}_2 >0$, incommensuration develops before $B$ changes sign,
and the transition into incommensurate SDW state remains second order over some
range of ${\delta}_0$ but eventually becomes first order 
at large ${\delta}_0$ and low $T$. The full phase diagram also contains a transition line
(not shown in \rfig{fig:sdwFFLO}) separating already developed commensurate and
incommensurate SDW orders. 

To analyze the interplay between the appearance of 
incommensurate SDW order and the sign change of $B$, 
we again expand the free energy in powers of $m$ but
now allow incommensuration parameter 
$\delta_1$ to be non-zero,
i.e., replace in the coefficients in \req{eq:GLfemag},
 ${\delta}_{\hvk} = {\delta}_0  + {\delta}_2 \cos(2\phi)$ with
${\delta}_{\hvk,\vq} = {\delta}_{\hvk} + {q} \cos(\phi - \phi_0)$. 
In general, for small ${q}$, 
\be 
\alpha_m({\delta}_{\hvk,\vq}) = \alpha_0 ({\delta}_{\hvk}) +  
\alpha_2 ({\delta}_{\hvk}) {q}^2 + \alpha_4({\delta}_{\hvk})  {q}^4 + {\cal O}({q}^6),
\label{ch_14}
\ee
with  $\alpha_0 ({\delta}_{\hvk})$  given by (\ref{eq:GLmag}).
When $\alpha_{4}$ and $B$ are positive, 
the N-SDW transition is second order, and is into a commensurate SDW state when
$\alpha_2 >0$ and into an incommensurate SDW state when $\alpha_2$
 changes sign and becomes negative.
If $B$ changes sign while $\alpha_2$ is still positive, the SDW-N transition
 becomes first order before incommensuration develops. 

To understand the phase diagram,  it is sufficient to consider small
${\delta}_2$. Expanding all coefficients in powers of ${\delta}_2$ we obtain 
\begin{subequations}
\label{ch_11}
\bea
\alpha_0 ({\delta}_{\hvk}) &=& 
\alpha_{0,0} + \alpha_{0,2} \; {\delta}_2^2 + \cO({\delta}_2^4)\,,
\\
\alpha_2 ({\delta}_{\hvk}) &=&  \alpha_{2,0} + 
\alpha_{2,1} \cos2\phi_0 \; {\delta}_2 + \cO({\delta}_2^2)\,,
\\
\alpha_4 ({\delta}_{\hvk}) &=&  \alpha_{4,0} + \cO(\delta^2_2), \nonumber \\
B &=&  \onehalf \alpha_{0,2} + \cO({\delta}^2_2) \,,
\eea
\end{subequations}
where
\begin{subequations}
\label{ch_12}
\bea
\alpha_{0,0} & = & \onehalf\left( \ln\frac{T}{T_s} + 2\pi T
 \sum_{{\vare}_n>0}
\frac{{\delta}_0^2}{{\vare}_n({\vare}_n^2+{\delta}_0^2)} \right) \,,
\\ 
\alpha_{2,0} & = & \alpha_{0,2} = \onefourth \, 2\pi {T} \sum_{{\vare}_n>0} {\vare}_n
\frac{{\vare}_n^2-3 {\delta}_0^2}{({\vare}_n^2+{\delta}_0^2)^3},
\\ 
\alpha_{2,1} & = &  \frac32 \, 2\pi {T} \sum_{{\vare}_n>0}
\frac{{\vare}_n ({\delta}_0^2-{\vare}_n^2){\delta}_0}{({\vare}_n^2+{\delta}_0^2)^4} \,,
\\
 \alpha_4 & = & -\frac{3}{16} \, 2\pi {T} \sum_{{\vare}_{n>0}}
{\vare}_n \frac{{\vare}_n^4-10 {\delta}_0^2 {\vare}_n^2+ 5{\delta}_0^4 }
{({\vare}_n^2+{\delta}_0^2)^5} \,.
\eea
\end{subequations}
We see from Eqs.~(\ref{ch_11})
that for  ${\delta}_2 =0$, $B$ and $\alpha_2 ({\delta}_{\hvk})$
change sign  simultaneously, at the point where 
$\alpha_{2,0}=\alpha_{0,2} =0$. 
However, when ${\delta}_2 \neq 0$, $\alpha_2 ({\delta}_{\hvk})$ changes sign
before $B$ becomes negative because $\alpha_2 (\delta_{\hvk})$ contains a term
linear in ${\delta}_2$, whose prefactor can be made negative by adjusting
$\phi_0$.  This explains why in \rfig{fig:sdwFFLO} incommensuration begins
while $B$ is still positive.  Also, we verified that near the onset points for
incommensuration, $\alpha_4 ({\delta}_{\hvk}) >0$, i.e., in this range
the transition into incommensurate SDW is second order. At larger
${\delta}_0$, the incommensurate transition eventually becomes first order.

\section{ SDW+SC state, numerical analysis } 
\label{sec:numer}

\begin{figure}[b]
\centerline{\includegraphics[width=\linewidth]{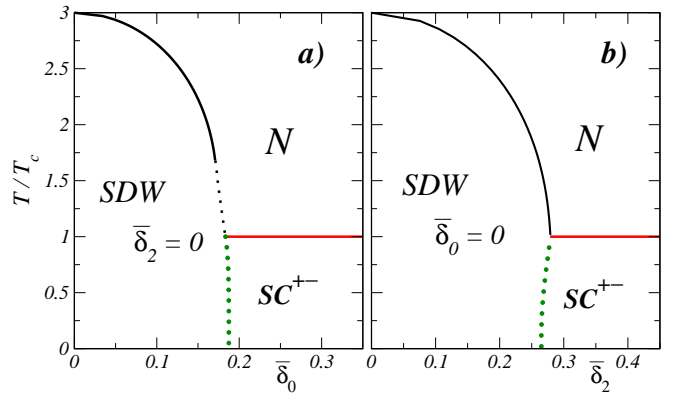}}
\caption{(Color online) 
	The phase diagram of SDW and SC  $s^\es$ states when only a commensurate
	SDW order is allowed ($q=0$).  We set  $T_s/T_c=3$ and varied either 
	the relative radius of circular hole and electron pockets (a)
	or  the form of one of the pockets (b). 
	The pure SC $s^\es$ and SDW states are separated by first
	order transition, and there is no co-existence region.
\cite{Vavilov09gl,Vorontsov09magsc} 
}
\label{fig:pdSPnocx}
\end{figure}

In the next two sections 
 we look at potential co-existence of SDW and the 
$s^\es$ or $s^\pp$ states, when the system is doped and 
the SDW state is suppressed. 
The  superconducting $T_c$ is  doping independent, so at some doping SDW and SC
transition temperatures cross.
Near this point, the two orders either 
support or suppress
each other and  either
co-exist or are separated by  a first-order transition.   

In  this section we present numerical results in the extended range of 
temperatures and dopings, in the next section we corroborate them with 
analytical consideration in the vicinity of the crossing point, 
when both order parameters are small, and at $T=0$.

\subsection{ Coexistence with $s^\pm$ state }

We look first at the $s^\es$ state.  In this case the system of coupled 
self-consistency equations for $\Delta$ and $m$ is obtained from 
\reqs{eq:scSCqc}-(\ref{eq:feqc}) by taking 
$ \Sigma^2_\pm = (E_n\pm i\delta_{\hvk \vq})^2 + m^2\, $
and
$E_n =  \sqrt{\vare_n^2 + \Delta^2 }$,
\begin{subequations}\label{eq:scboth}
\be
\ln \frac{T}{T_c} = 2 \pi  T \sum_{\vare_n >0 }
\Re \left\langle \frac{ (E_n+i\delta_{\hvk \vq})/E_n }{\sqrt{(E_n+i\delta_{\hvk \vq})^2+m^2} }
-\frac{1}{|\vare_n|} \right\rangle \,,
\label{eq:scSCt}
\ee
\be
\ln \frac{T}{T_s} = 2 \pi T \sum_{\vare_n > 0}\;
\Re \left\langle \frac{ 1}{ \sqrt{(E_n+i\delta_{\hvk \vq})^2+m^2} }
-\frac{1}{|\vare_n|} \right\rangle \,.
\label{eq:scSDWt}
\ee
\end{subequations}
We remind that  $T_c$ is the transition temperature for the pure SC state,
and $T_s$ is the transition temperature for the 
pure SDW state at $\delta_{\hvk \vq} =0$.

\begin{figure}[t]
\centerline{\includegraphics[width=0.95\linewidth]{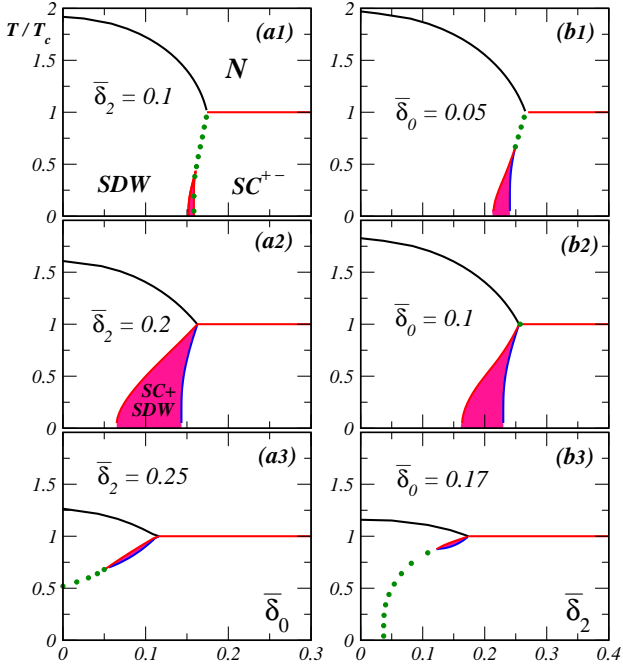}}
\caption{(Color online) 
	Appearance of co-existence when both $\delta_0$ and $\delta_2$ are finite. 
	We set $T_s/T_c=2$ and $q=0$. Panels (a1)-(a3) -- phase diagrams in
	variables $T,\delta_0$  at fixed $\delta_2$, panels (b1)-(b3) -- phase
	diagrams in 
	variables $T,\delta_2$ at fixed $\delta_0$.  Panels (a1), (b1) -- there
	appears a region near $T=0$, where SDW and SC $s^{+-}$ orders co-exist.
	Panels (a2),(b2) -- the co-existence region broadens and reaches $T=T_c$.
	Panels (a3), (b3) -- the transition at low $T$ becomes first order
	between pure SDW and SC states, but narrow co-existence region is still present
	near $T_c$.  A complimented  zero-temperature phase diagram is presented in 
	Fig.\protect\ref{fig:pdSPzeroT}.
}
\label{fig:tsc2pd}
\end{figure}

\begin{figure}[t]
\centerline{\includegraphics[width=\linewidth]{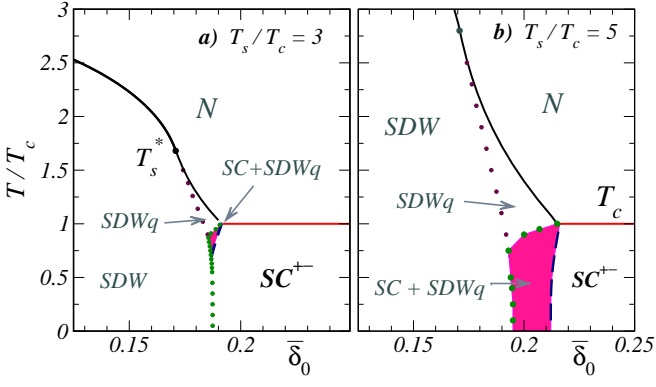}}
\caption{(Color online) 
	The phase diagram for $\delta_2 =0$, 
	when the system can choose the value of $\vq$.
	Incommensurate SDW order appears below $T_s^*=0.56 T_s$ 
	and leaves some parts of the FS ungapped, allowing for co-existing SC order. 
		(a) $T_s/T_c =3$. 
	The 
	SDW+SC phase appears only in a small region near $T_c$. At 
	low $T$ the system still undergoes a first order transition 
	between commensurate SDW and  SC states. 
	(b) For larger $T_s/T_c =5$ (weaker SC interaction) 
	the co-existence region widens and extends down to $T=0$.
	The $q=0$ 
	SDW state 
	has the lowest energy at $T=0$ for ${\bar\delta}_0 \lesssim 0.195$. 
	}
\label{fig:cx35}
\end{figure}

These equations are solved 
numerically to find all possible states $(\Delta, m)$ 
and their energies evaluated using \req{eq:feqc}. 
The main results for this part are presented in  
Figs.~\ref{fig:pdSPnocx}-\ref{fig:FFLO}. 

\subsubsection{Commensurate SDW state}

Figure \ref{fig:pdSPnocx} shows the results for the case when SDW order is set
to be commensurate (i.e., $q =0$) and the FSs are either co-axial
circles (panel a), or of different shapes with equal $k_F$ (panel b). In
the first case, ${\delta}_2 =0$ and ${\delta}_0 \neq 0$, in the second case
${\delta}_0 =0$ and ${\delta}_2 \neq 0$.
We see that in both cases pure SDW and
SC states are separated by a first-order transition. 
We verified  that
in both cases fermionic excitations in the SDW state are fully
gapped at $T=0$ and thus there are no Fermi surfaces.  From this perspective, the
results presented in Figure \ref{fig:pdSPnocx} are consistent with the idea
that co-existence requires the presence of the Fermi surfaces in the SDW state.
However, we will see next that the situation in the cases when both $\delta_0$
and $\delta_2$ are non-zero is more complex.  

This is demonstrated in \rfig{fig:tsc2pd} which shows the phase diagram for
$T_s/T_c=2$ as a function of $\delta_0$ for a set of fixed $\delta_2$ (panels
(a1)-(a3)), and as a function of $\delta_2$ for a set of fixed $\delta_0$
(panels (b1)-(b3)).  For all cases, pure SDW state is fully gapped at $T=0$, so
naively one should not expect a co-existence state. However, as is evident from
the figure, the phase diagram does involve the co-existence phase, which can be
either at low $T$ (including $T=0$), or near $T=T_c$, depending on the
parameters. In particular, as $\delta_2$ in panels (a) or $\delta_0$ in panels
(b) increase, the co-existence state first appears at low $T$, while at higher
$T$ the pure SDW and SC states are still separated by first-order transition
(panels (a1) and (b1)). Then the co-existence region grows, and extends up to
$T=T_c$ (panels (a2) and (b2)). At even larger $\delta_2$ or $\delta_0$, SDW
and SC states are separated by the first-order transition at low $T$, but the
co-existence phase still survives near $T_c$. 

\begin{figure}[t]
\centerline{\includegraphics[width=\linewidth]{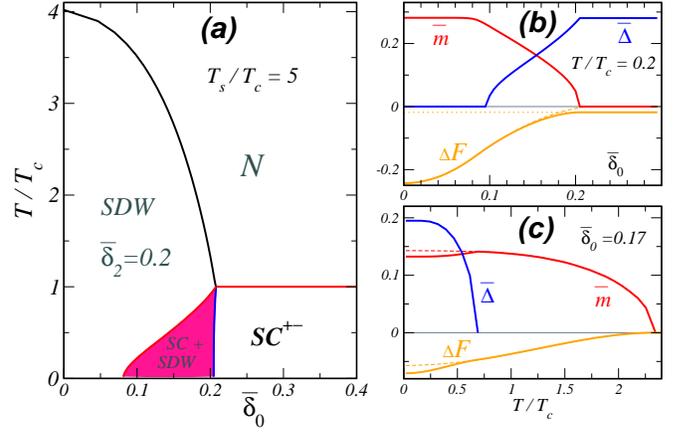}}
\caption{(Color online) 
	(a) Same as in \rfig{fig:tsc2pd}(b) but for $T_s/T_c=5$. 
	(b,c) SDW and SC gaps, in 
	units ${\bar m}$ and ${\bar \Delta} = \Delta/2\pi T_c$,
	and the free energy 
	as functions of ${\bar\delta}_0$ along the line $T/T_c=0.2$ (b), 
	and as functions of temperature for ${\bar\delta}_0 = 0.17$ (c). 
}
\label{fig:ellipscx}
\end{figure}
\begin{figure}[t]
\centerline{\includegraphics[width=0.9\linewidth]{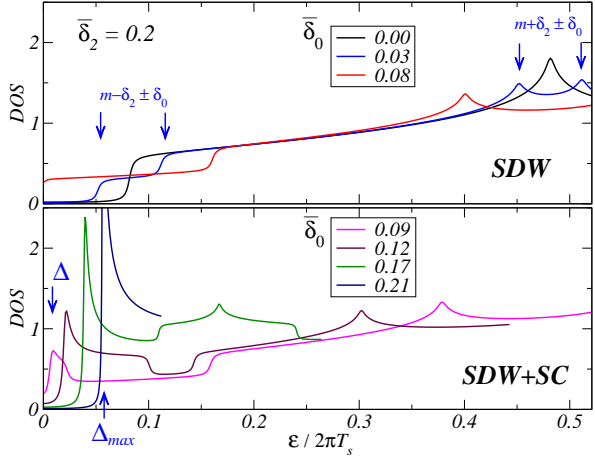}}
\caption{(Color online) 
	FS averaged DOS as a function of energy at $T / T_c =0.1$ for different
	values of ${\delta}_0$ (i.e., different dopings).  
	We set ${\bar \delta}_2 = 0.2$ and  $T_s/T_c = 5$. 
	The characteristic values of the SDW order parameter for this range of parameters 
	is ${\bar m} \approx 0.28$ (zero-$T$ limit). 
	Upper panel: DOS for
	small ${\bar \delta}_0$, when the system remains in the pure SDW state.
	This figure is similar to \rfig{fig:dosSDW}. 
	The DOS vanishes below ${\epsilon} = {m} - {\delta}_0 - {\delta}_2$ 
	and has $\ln-$ non-analytic behavior
	at ${\epsilon} = {m} \pm {\delta}_0 + {\delta}_2$, 
	and sudden drops at ${\epsilon} = {m} \pm {\delta}_0 - {\delta}_2$. 
	Lower panel: DOS for
	larger ${\delta}_0$, when SDW and SC orders co-exist. Sharp peaks at
	small ${\epsilon}$ are due to opening of the superconducting gap $\Delta$. 
	Once SDW order disappears at $\bar\delta_0 \approx 0.21$, the DOS
	acquires BCS form with the maximal gap $\Delta_{max}/2\pi T_c \approx 0.28$. 
	} 
	\label{fig:dosSDWSC} 
\end{figure}
In Fig. \ref{fig:ellipscx} we show the phase diagram for ${\bar \delta}_2 =
0.2$ and $T_s/T_c=5$ together with the plots of SDW and SC order parameters and
the free energy. We see the same behavior as in \rfig{fig:tsc2pd} (a2)  --
there is a co-existence phase for all $T$ up to $T_c$.  In \rfig{fig:dosSDWSC}
we show the changes in the quasiparticle DOS at low $T =0.1 T_c$ as the system
evolves  from the SDW state to the SC state  via the co-existence region. 

\begin{figure}[t]
\centerline{\includegraphics[width=\linewidth]{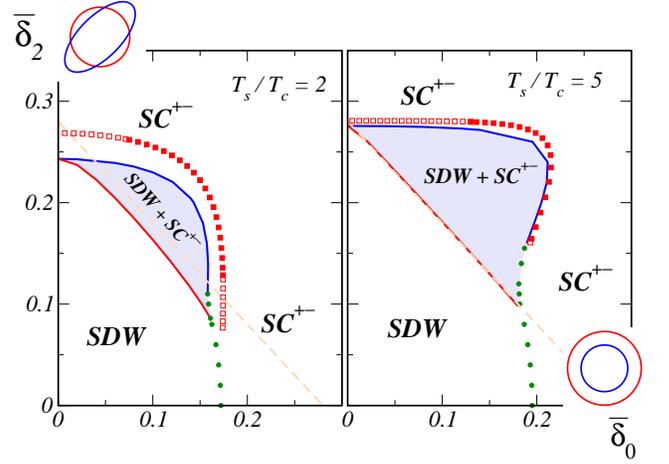}}
\caption{(Color online) 
	The zero temperature phase diagram: SDW, SC states and their co-existence
	region for various $\delta_2$ and $\delta_0$ and two different $T_s/T_c =2$
	(left) and $T_s/T_c =5$ (right). We only allow the system to
	develop a commensurate SDW order ($q =0$).  
	The dashed line denotes first appearance of gapless excitations in the pure SDW state
	[$m = \delta_0 + \delta_2$, \protect\req{eq:dosSDW}]. 
      The co-existence region at $T=0$ (shaded area) 
	extends down to $\delta_0=0$, but is not
	present at small $\delta_2$. The width of the co-existence region increases
	with the relative strength of SDW interaction, as determined by ratio $T_s/T_c$.  
	The squares mark the location of the crossing point between 
	$T_s (\delta_0,\delta_2)$ and $T_c$. 
	Open squares indicate that the SDW-SC
	transition near $T_c$ is first order, while filled squares 
      signal the presence of the co-existence phase near $T_c$. Note that the regions
	where co-existence phase is present at $T=0$ and near $T_c$ are not
	identical.
} 
\label{fig:pdSPzeroT} 
\end{figure}

Finally, in \rfig{fig:pdSPzeroT}, we show  the zero-temperature phase diagram
in variables ${\delta}_0$ and ${\delta}_2$ for $T_s/T_c =2$ and  $T_s/T_c =5$,
together with the locus of points where $T_s (\delta_0,\delta_2) = T_c$.  The
phase diagram was obtained by numerically solving \reqs{eq:scboth} and
evaluating the free energy at $T/T_c = 0.02$.  This phase diagram corroborates
the results of \rfig{fig:tsc2pd} and \ref{fig:ellipscx} --  the zero T
behavior in panels (a) and (b) in  \rfig{fig:tsc2pd}) is obtained by taking
either horizontal or vertical cuts in \rfig{fig:pdSPzeroT}. In particular, we
see from \rfig{fig:pdSPzeroT}, that the transformation between  panels (a2) and
(a3)  of \rfig{fig:tsc2pd} is such that the co-existence region at $T=0$ first
moves to the left, shrinks, and disappears at ${\bar \delta}_2 \approx 0.24$.
Similarly, in panels (b), the co-existence range shrinks to a point at ${\bar
\delta}_0 \approx 0.16$, and at larger $\delta_0$ the transition between SDW
and SC phases at $T=0$ becomes first order.  In the next section we present the
results of complimentary analytical studies of the phase diagram at $T=0$ and
near $T_c$. These results are in full agreement with the numerical analysis in
this section.

Observe that for $T_s/T_c =5$ the left boundary of the co-existence region is
located very close to the line $\delta_0 + \delta_2 = m_0$ (dashed line in
\rfig{fig:pdSPzeroT}, ${\bar \delta}_0 + {\bar \delta}_2 = 0.28073$), at which
gapless excitations and Fermi surfaces appear in the SDW state. For this
$T_s/T_c$, the co-existence region at $T=0$ virtually coincides with the region
where SDW state has a Fermi surface.  However, for smaller $T_s/T_c =2$ 
(\rfig{fig:tsc2pd}; left panel in \rfig{fig:pdSPzeroT}), 
co-existence clearly occurs already in the
parameter range where SDW excitations are all gapped. The co-existence for 
$T_s/T_c =2$ is therefore not the result of the ``competition for the Fermi
surface'', but rather the consequence of the fact that the system can gain in
energy by reducing 
the SDW order parameter (still keeping all
fermionic excitations gapped) and creating a non-zero SC order parameter. The
gain of energy in this situation can best be interpreted as the consequence of
the attraction between the two orders.

\begin{figure}[t]
\centerline{\includegraphics[width=0.6\linewidth]{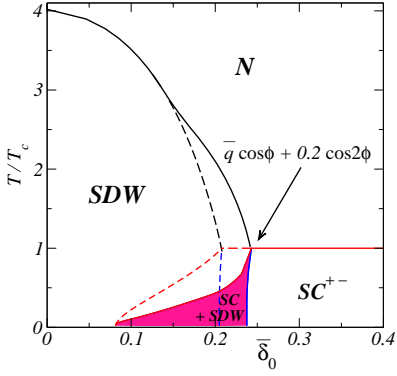}}
\caption{ (color online)
	Same as in \rfig{fig:ellipscx}(a) but now we allow the system to choose 
	the value of ${q}$. The phase diagram from \rfig{fig:ellipscx}
	is shown by dashed lines.
	A finite ${q}$ emerges below a particular $T$ and moves 
	the co-existence region to 
	larger ${\delta}_0$, 
	together with the SDW-N transition. This  broadens 
      the co-existence region, 
	and slightly changes the shape of 
	$T_c(\delta_0)$ inside the magnetic dome.
	}
\label{fig:FFLO}
\end{figure}

\subsubsection{Commensurate vs. incommensurate SDW state}

One of the results of our consideration so far is that, if we keep   
an SDW order commensurate, a finite region of  SDW + SC phase appears only when
both ${\delta}_0$ and ${\delta}_2$ are non-zero. If we allow the system to
choose  the ordering momentum of the SDW state, the co-existence region widens
and appears even if we set $\delta_2=0$. We illustrate this in \rfig{fig:cx35},
where we plot the phase diagram at $\delta_2=0$ for two different values of
$T_s/T_c$.  In agreement with Fig. \ref{fig:sdwFFLO},  at $T <T^*_s$, the
system chooses an SDW state with a non-zero ${q}$. We see that,  in this
situation, there appears a region where SC state co-exist with an
incommensurate SDW state.\cite{Vorontsov09magsc} The co-existence region widens
up when the ratio $T_s/T_c$ increases, and for large enough $T_s/T_c$ extends
down to $T=0$.  In \rfig{fig:FFLO} we set $\delta_2$ to be non-zero (${\bar
\delta}_2=0.2$) and allowed the system to choose ${q}$ which minimizes the free
energy.  The results  are quite similar to the case when $q=0$.  We see that
the SDW and SC orders do co-exists in the parameter range which extends from
the crossing point down to $T=0$. The width of the co-existence region widens a
bit when we allow the system to choose ${q}$, but qualitatively, the behavior
in Figs.~\ref{fig:ellipscx} and \ref{fig:FFLO} is the same.  Note, in our
two-band model, the ellipticity of of the electron FS breaks the rotational
symmetry and favors the direction of $\vq$ along the ellipse's major axis, see
Eq.~(\ref{ch_11}b).

To summarize, SDW and SC$^{+-}$ phases do co-exist in a range of finite
dopings, but the width of the co-existence region depends on the amount of
ellipticity of the electron band and the ratio of $T_s/T_c$. At larger
$T_s/T_c$  the width of the co-existence region increases for fixed 
${\delta}_2$, and there is optimal ${\delta}_2$ at which the width is the
largest.  The fact that the system can lower the energy by making SDW order
incommensurate also acts in favor of co-existence,
but qualitatively the picture remains the same as in the case when $q$
is set to be zero. 

\subsection{ Minimal co-existence with $s^\pp$ state } 

We next look at the SC state with gaps of the same signs on two FSs. 
Such states seem unlikely for pnictides,   because
they require a negative sign of the interband pair hopping term.\cite{Chubukov08}
Still, it would be interesting to investigate consequences of 
attractive SC interaction between electron and hole bands. 

\begin{figure}[b]
\centerline{\includegraphics[width=0.9\linewidth]{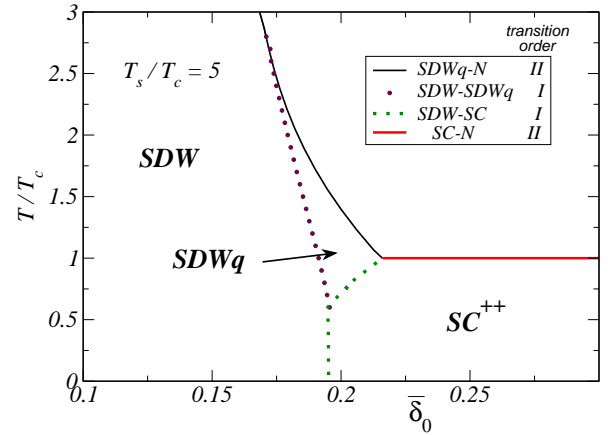}}
\caption{(Color online) 
	The phase diagram for a conventional $s^\pp$ SC order parameter, at
	${\delta}_2 =0$, and varying ${\delta}_0$. We allow the system to
	choose ${q}$.  
	The SDW+SC state does not appear, even when SDW order becomes
	incommensurate.
}
\label{fig:SSpd}
\end{figure}

The expressions for $\Sigma_\pm$ in this case is slightly 
more complicated 
and less illuminating than those for $s^\es$ state, although 
quite similar, and so are the self-consistency equations, 
which we do not write here, but which are obtained from Eqs.~(\ref{eq:scSCqc})-(\ref{eq:feqc})
in a way completely analogous to \reqs{eq:scboth}. 
We first present the results for ${ \delta}_2=0$, \rfig{fig:SSpd}. 
We found that co-existence
region does not appear even if we allow SDW order to become incommensurate.  
There are commensurate and incommensurate
SDW phases on the phase diagram, and SC$^{++}$ phase, but the transition
between SC and SDW phases remains first order.   In other words, the appearance
of gapless excitations in the SDW phase due to incommensuration at large
$\delta_0$ does not seem to favor a mixed superconducting and magnetic state,
in sharp contrast to the case of  $s^\es$ SC, where incommensuration induces
co-existence, see \rfig{fig:cx35}(b). 

\begin{figure}[t]
\centerline{\includegraphics[width=0.9\linewidth]{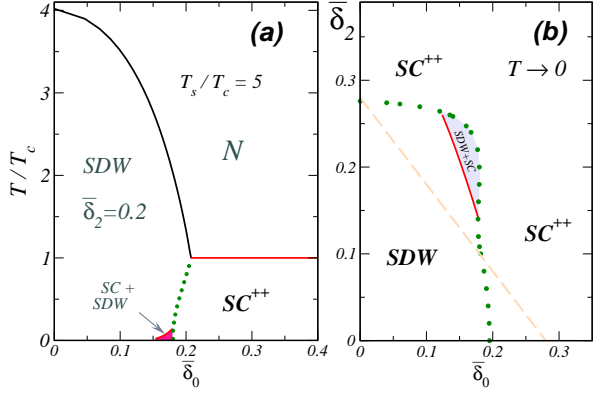}}
\caption{(Color online) 
(a) Same as in \rfig{fig:SSpd} but for fixed ${\bar \delta}_2=0.2$ and 
${q} =0$. For most of the phase diagram, the behavior is the same as for
${\delta}_2 =0$,  
but there appears a very tiny range of SC + SDW phase at the lowest $T$.
The transition to purely SC state is first order at all $T$. 
(b) The zero-temperature phase diagram.
SDW and SC $s^\pp$ states are separated by first order transition virtually
everywhere except a small region at finite ${\delta}_0$ and ${\delta}_2$,
where SDW+SC state emerges. }
\label{fig:SSzeroT}
\end{figure}

For a non-zero  ${ \delta}_2$, 
there might appear a tiny region of co-existence
at low temperatures.  We illustrate this in \rfig{fig:SSzeroT}, where in
panel (a) we plot the phase diagram for 
${\bar \delta}_2 =0.2$ and set ${ q} =0$. 
(When the system is allowed to choose ${ q}$, 
the results change minimally, in a way similar to \rfig{fig:FFLO}). 
In panel (b) of this figure we show where the region 
of SDW+SC$^\pp$ exists for different ${ \delta}_0$ and ${\delta}_2$. 
 We see that the range of co-existence is very  narrow, and we
also found that the difference in free energies between a pure SDW state and
SDW+SC state is very small due to small value of the SC order parameter. 

Observe also that the co-existence region in \rfig{fig:SSzeroT} is to the left
of the line $\delta_0 + \delta_2 = m_0$   at which a Fermi surface appears in
the  SDW state (a dashed line in \rfig{fig:SSzeroT}b). In other words,
$s^{++}$ superconductivity does not emerge even when there is a Fermi surface
in the SDW state. This shows once again that the presence or absence of the
Fermi surface in the SDW state is not the primary reason for the presence or
absence of the SDW+SC phase. The true reason is energetic -- the SDW+SC state
can either lower or increase the energy compared to pure state depending on
whether SDW and SC orders attract or repel each other. The absence of the
co-existence phase even in the range where SDW state has a  Fermi surface is a
clear indication that there is the ``repulsion'' between SDW and SC
orders, if the SC order is  $s^\pp$, \reqs{eq:scSCqc}-(\ref{eq:scSDWc}). 
The same conclusion was recently reached by Fernandes {\it et al}.\cite{Fernandes09} 

\section{\label{sec:GL} SDW + SC, analytical results}

We corroborate the numerical analysis in the preceding Section with the
analytical analysis.  We first present the results of Ginzburg-Landau (GL)
description near the point where second-order SDW-N and SC-N transitions meet,
then consider the phase diagram at $T=0$, and finally combine the  two sets of
results and compare analytical phase diagram with Fig. \ref{fig:tsc2pd}.

\subsection{Ginzburg-Landau analysis}

We begin with the GL  analysis near the point where $T_s (\delta_0,\delta_2) =
T_c$.  Near this point, both the SDW and SC order parameters are small and we
can  expand the free energy, \req{eq:feqc}, to the fourth order in $m$ and
$\Delta$ and compare different phases. For simplicity, in this section we
assume that the SDW order is commensurate. An extension to a finite ${ q}$
complicates the formulas but does not change the outcome.

The expansion of the free energy, Eq. (\ref{eq:feqc}) in powers of $m$ and
$\Delta$ yields
\be 
\cF =\alpha_\Delta { \Delta}^2+\alpha_m { m}^2+A{ \Delta}^4+B
{ m}^4+2C{ \Delta}^2 { m}^2 \,.  
\label{eq:F} 
\ee
where $\cF = \Delta F (m,\Delta)/(4N_F)$.
 Coefficients $\alpha_\Delta$, $\alpha_m$, $A$, and 
$B$ in \req{eq:F} are identical for both
$s^\es$ and $s^{\pp}$ SC states:
\bea
\alpha_\Delta & = & \onehalf \ln\frac{T}{T_c} \,,  
\\
\alpha_m & = & \onehalf \left(  \ln\frac{T}{T_s}
+2\pi { T}\sum_{{ { \vare}}_n>0}\left\langle
\frac{{ \delta}_{\hvk}^2}{{ \vare}_n({ \vare}_n^2+{ \delta}_{\hvk}^2)}\right\rangle 
\right) \,,
\label{ch_7}
\eea
and 
\bea 
A& = &  \frac{ \pi { T}}{4}\sum_{{ \vare}_n>0}\frac{1}{{ \vare}_n^3},
\label{eq:A}\\
B&  =&  \frac{\pi { T}}{4}\sum_{{ \vare}_n>0}\left\langle { \vare}_n
\frac{{ \vare}_n^2-3 { \delta}_{\hvk}^2}{({ \vare}_n^2+{ \delta}_{\hvk}^2)^3}\right\rangle.
\label{eq:B}
\eea
The difference between $s^{\es}$ and $s^{\pp}$ SC orders
appears only in the coefficient $C$. For $s^{\es}$ state we have
\be
C_{(\es)}= \frac{\pi { T}}{4}\sum_{{ \vare}_n>0}\left\langle 
\frac{{ \vare}_n^2-  { \delta}_{\hvk}^2}
{{ \vare}_n({ \vare}_n^2+{ \delta}_{\hvk}^2)^2}\right\rangle\,,
\label{eq:C}
\ee
while for $s^{\pp}$
\be
C_{(\pp)} = \frac{\pi { T}}{4}\sum_{{ \vare}_n>0}\left\langle 
\frac{3 { \vare}_n^2+  { \delta}_{\hvk}^2}{{ \vare}_n({ \vare}_n^2+{ \delta}_{\hvk}^2)^2}
\right\rangle\,.
\label{eq:Cpp}
\ee

Note, that, although both $C$-coefficients are positive, this does not 
preclude co-existence in \req{eq:F}, and we find below that 
the sign of parameter $\chi=AB-C^2$ is more important for co-existence. 
We will demonstrate that since $C_{(\pp)} > C_{(\es)}$,  
$\chi$ is positive for a broader range of parameters in $s^\es$ state
than that in  $s^\pp$ state. In fact, $\chi$ remains always negative in $s^\pp$ state.  
Below we will use the 
notion that $\chi >0$ corresponds to an effective
attraction between the two orders.

The free energy, \req{eq:F}, has two local minima, corresponding to pure
states, when one of the order parameters is identically equal to zero: 
\\
1) a pure SC state, defined by ${ m}=0$ and $\partial \cF/\partial { \Delta}=0$, has
the free energy and SC order parameter
\be
\label{eq:Fs}
\cF_\Delta = -\frac{\alpha_\Delta^2}{4A}, \quad { \Delta}^2 = -\frac{\alpha_{ \Delta}}{2A};
\ee
2)  a pure SDW state, defined by  
${ \Delta}=0$ and $\partial \cF/\partial { m}=0$, has the free energy and SDW
order parameter 
\be
\label{eq:Fm}
\cF_{ m}  = -\frac{\alpha_m^2}{4B},\quad { m}^2 = -\frac{\alpha_m}{2B}.
\ee

In addition, the free energy  may also 
have either a saddle point or a global minimum when both
${ \Delta}\neq 0$ and ${ m}\neq 0$.
To see this, we write the free energy \req{eq:F} in equivalent form, 
\bea
\cF
= && \alpha_m \left( { m}^2 + \frac{C}{B}{ \Delta}^2 \right) 
     + B \left( { m}^2 + \frac{C}{B}{ \Delta}^2 \right)^2 
    \nonumber \\ \nonumber 
     &&+ \left( \alpha_\Delta - \frac{C}{B} \alpha_m \right) { \Delta}^2 
     + \left( A - \frac{C^2}{B} \right) { \Delta}^4 \,,
\eea
which is now a sum of two independent parts 
for ${ \Delta}^2$ and ${ M}^2 \equiv { m}^2 +(C/B){ \Delta}^2$.
For an extremum state, given by $\partial_{ \Delta}\cF=\partial_{ m}\cF=0$, 
the stationary values of order parameters,  
\be
{ \Delta}^2 = -\frac{\alpha_\Delta B - \alpha_m C}{ 2 (AB - C^2) }
\quad,\quad
{ M}^2 =  { m}^2 + \frac{C}{B}{ \Delta}^2 = -\frac{\alpha_m}{2B} \,,
\label{starstate}
\ee
determine the free energy, 
 \be
\cF_{m\& \Delta} = -B { M}^4 - \frac{AB-C^2}{B} { \Delta}^4 \,.
\label{Fquadratic}
\ee
When both coefficients in \req{Fquadratic} are positive, 
\be
B >0  \quad, \qquad \chi=AB - C^2 >0 \,,
\label{aa}
\ee
the mixed state, \req{starstate}, corresponds to the minimum of the free energy,
which is smaller than the minima for pure SC or SDW states:
\be
 \label{eq:Fmix}
 \begin{split}
   \cF_{m\&\Delta}
   & = \cF_m-\frac{1}{4B}\frac{(\alpha_\Delta B -\alpha_m C)^2}{AB-C^2} \\
   & =  \cF_\Delta-\frac{1}{4A}\frac{(\alpha_m A-\alpha_{ \Delta} C)^2}{AB-C^2}.
 \end{split}
\ee
Consequently, in the phase diagram, the pure SDW and SC states are separated by
a SDW+SC phase, and the transitions into this intermediate state are
second-order.  However, if $B >0$ and $\chi <0$, the mixed phase,
\req{starstate}, corresponds to the saddle point of the free energy and is not
thermodynamically stable phase.  In this case, pure SDW and SC phases are
separated by a first-order transition line.  When $B<0$, one needs to expand
further in $m$ to determine the phase diagram. We will not discuss the case $B
<0$ further within GL theory.

We apply \req{aa} to the case  ${ \delta}_{\hvk} = { \delta}_0 + { \delta}_2
\cos 2 \phi$ which we considered in the previous Sections.  We remind that ${
\delta}_2 =0$ corresponds to co-circular FSs with different chemical
potentials, while ${ \delta}_0=0$ corresponds to FS geometry in which $k^c_F =
k^f_F$,  but the electron pocket is elliptical. 

At perfect nesting ${ \delta}_{0} = { \delta}_2 =0$,
and the system develops an SDW order at $T_s > T_c$.
Deviations from perfect nesting lead to two effects. First, as we already said,
the magnitude of  $\alpha_m$ is reduced  because SDW instability is suppressed
when nesting becomes non-perfect.  Superconducting $\alpha_\Delta$ is not
affected by  ${ \delta}_{\hvk}$, and eventually 
wins over SDW.  Second, coefficients $B$ and $C$ evolve with
${ \delta}_{\hvk}$ and, as a result, the sign of  $\chi =AB-C^2$ depends on
values of ${ \delta}_{0}$ and ${ \delta}_{2}$.

The GL expansion is applicable only in
the vicinity of points at which the temperatures  of the SDW-N and SC-N
transitions coincide $T_s(\delta_{\hvk})=T_c$. This condition together with 
\req{eq:ts}  establish the 
 relation between 
 ${ \delta}_0$ and ${ \delta}_2$ at 
 which one needs to compute the parameters $B$ and $C$. 

\subsubsection{$s^{+-}$ superconductivity}

To get an insight on how $\chi$ evolves with $\delta_{\hvk \vq}$, we
first assume that $T_s/T_c$ is only slightly larger than one 
($T_s/T_c = 1 + \delt$), in which case 
$T_s(\delta_{\hvk})=T_c$ at small $\delta_0$ and $\delta_2$, and we can 
expand $A$, $B$, and $C$ in powers of ${ \delta}_0$ and ${ \delta}_2$. 
 Specifically, we have from \req{eq:ts}
\be
\delt = \frac{7\zeta (3)}{4\pi^2 T^2_s} \left(\delta^2_0 + \frac{1}{2}
\delta^2_2\right) = \frac{0.663}{m^2_0}  \left(\delta^2_0 + \frac{1}{2}
\delta^2_2\right) 
\label{dr_1}
\ee
where $\zeta (3)$ is a Riemann Zeta function.
 Collecting terms up to the fourth order in the expansion, we obtain 
\be
\chi=\frac{1}{32\pi^8{ T}_c^8}\left(
s_1\langle{ \delta}_{\hvk}^4\rangle-s_2 \langle{ \delta}_{\hvk}^2\rangle^2
\right),
\ee
where
\bea
&&s_1 = 5 \left(\sum_{n\ge0} \frac{1}{(2n+1)^3} \right)\left(  \sum_{n\ge0}
\frac{1}{(2n+1)^7} \right),
\nonumber\\
&&s_2 = 9 \left(\sum_{n\ge0} \frac{1}{(2n+1)^5}\right)^2.  
\eea
The sums are expressed in terms of the Riemann-Zeta function 
 $\zeta(3), \zeta(5)$, and $\zeta(7)$ and 
give  $s_1\approx 5.261$ and $s_2\approx 9.082$.
Substituting ${ \delta}_{\hvk}={ \delta}_0+{ \delta}_2\cos 2\phi$ 
and averaging over
momentum direction $\phi$ on the FSs, we obtain
\be
\chi\approx \frac{1}{32\pi^8{ T}_c^8}\left(
-3.820{ \delta}^4_0 +6.703 { \delta}^2_0 { \delta}^2_2 - 0.297 { \delta}^4_2\right).
\label{eq:chi_exp}
\ee
\begin{figure}[t]
\centerline{\includegraphics[width=1.0\linewidth]{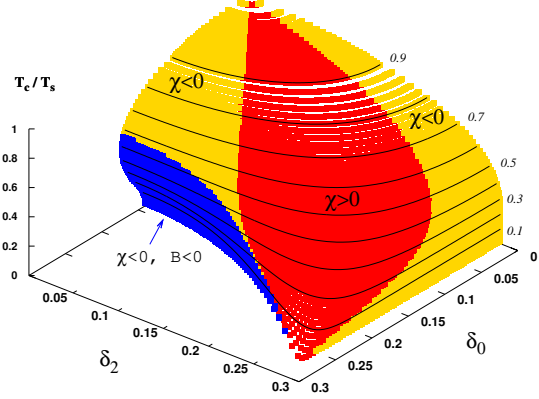}}
\caption{(Color online)
	The three-dimensional plot of the SDW-SC crossing surface, 
	$T_s({ \delta}_0, { \delta}_2) = T_c$.
	At each point on the surface we show the sign of $B$ and $\chi = AB - C^2$.  
	In the region where $\chi > 0$, the transition between 
	SDW and SC states occurs via the
	co-existence region. For $\chi<0$, pure SDW and SC 
	states are separated by the first-order transition.    
	When $B<0$, the SDW-N conversion is of the first order and
	the present GL analysis is invalid. 
	}
\label{fig:GL}
\end{figure}
We see that for ${ \delta}_0={ \delta}_2 =0$, $\chi =0$, 
i.e., for a perfect nesting
the system cannot distinguish between first order transition and SDW+SC phase.
This result, first noticed in Ref. \onlinecite{Fernandes09}, implies that the phase
diagram is quite sensitive to the interplay between ${ \delta}_0$ and ${ \delta}_2$.   
We see from (\ref{eq:chi_exp}) that in the two limits when
either ${ \delta}_2 =0$ or ${ \delta}_0=0$, $\chi <0$, i.e., the transition is first
order. This agrees with the numerical analysis in the previous Section. 
We emphasize that in  both limits, a small SDW order, which we consider here, 
still preserves low-energy fermionic states near the modified FSs. Fermions near
these FSs do have a possibility to pair into  $s^{\es}$ state. 
However, SDW+SC state turns out to be energetically unfavorable. 
We particularly emphasize that the ellipticity of
electron dispersion is not sufficient for the appearance 
of the SDW+SC phase near  $T_c \sim T_s$. 
 
When both ${ \delta}_0\neq 0$ {\it and} ${ \delta}_2\neq 0$, there is a
broad range 
\be\label{eq:delta2range}
0.765<\frac{\delta_2}{\delta_0}<4.689,
\ee 
where $\chi >0$ and the
transition from a pure SDW phase to pure a SC phase
occurs via an intermediate phase where the two orders co-exist.  This also
agrees with the numerical analysis (see \rfig{fig:ellipscx} and 
\rfig{fig:pdSPzeroT}). 

 Eq. (\ref{eq:delta2range}) has to be combined with the equation for $T_s
(\delta_0,\delta_2) =T_c$, and the boundaries in Eq. (\ref{eq:delta2range}) set
the critical values of $\delta_2$ and $\delta_0$ as functions of $T_s/T_c$.
Combining \reqs{eq:delta2range} and (\ref{dr_1}), we obtain that
co-existence occurs for 
\be 0.826 m_0 \sqrt{\delt} < \delta_2 < 1.663 m_0
\sqrt{\delt} \label{dr_5} 
\ee   

To verify that this result holds at larger values of ${ \delta}_0$ and ${ \delta}_2$,
we computed $\chi$ without expanding in $\delta_{\hvk \vq}$. We plot the
resulting phase diagram in \rfig{fig:GL}. 
The result is qualitatively the
same as \req{eq:chi_exp}: for ${ \delta}_0 =0$ or ${ \delta}_2 =0$, $\chi <0$
and the transition between SDW and SC states is of first order, while when both
${\delta}_0$ and ${ \delta}_2$ are non-zero, there exists a region where $\chi >0$
and the transition from SDW to SC state occurs via an intermediate SDW+SC
phase.

\subsubsection{$s^\pp$ superconductivity}

\begin{figure}[t]
\centerline{\includegraphics[width=0.90\linewidth]{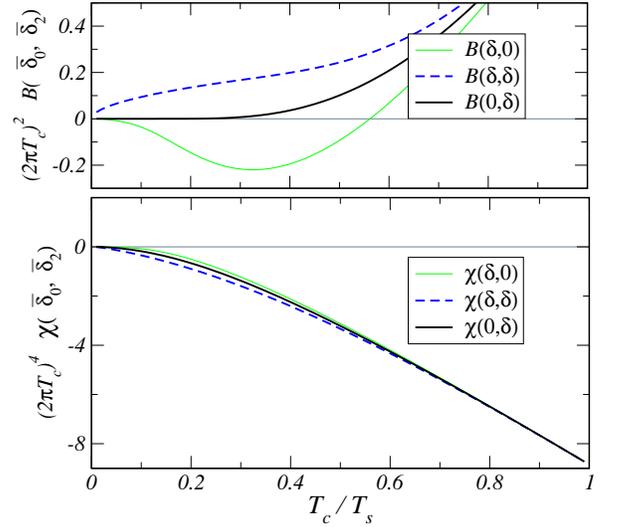}}
\caption{(Color online)  
The case of $s^{++}$ SC. Panel (a): 
the  behavior of $\chi({ \delta}_0,{ \delta}_2)=AB-C^2$ 
for different $T_s/T_c$ for three cases:
$\chi(\delta,0)$, $\chi(0,\delta)$, and $\chi(\delta,\delta)$. 
For each case, $\delta$ is chosen to satisfy  the condition
 $T_s({ \delta}_0, { \delta}_2)=T_c$ for a given $T_s/T_c$. We see that in all
 three cases, $\chi <0$ no matter what the ration $T_s/T_c$ is. 
Panel B -- the coefficient $B ({ \delta}_0,{ \delta}_2)$ along the line
  $T_s({ \delta}_0, { \delta}_2)=T_c$. GL analysis is only valid when $B >0$. }
\label{fig:GLpp}
\end{figure}

We performed the same calculations for a conventional, sign-preserving $s$-wave
superconductivity. The key difference with the $s^{+-}$ case is that now 
$\chi = AB-C^2$ is non-zero already when ${ \delta}_0={ \delta}_2=0$.  Substituting $A$ and
$B$ from (\ref{eq:A}) and (\ref{eq:B}) and $C$ from (\ref{eq:Cpp}) we obtain 
\be
\chi (\delta_{\hvk \vq} =0) = - \frac{7 \zeta (3)}{128 \pi^4 T_c^4} <0
\ee
The implication is that, for small ${ \delta}_0$ and ${ \delta}_2$, $\chi$ remains
negative and the transition between SDW and SC states is first-order. 
This result was  first obtained  by Fernandes \et in
Ref.~\onlinecite{Fernandes09}. These authors also argued,  based on their
numerical analysis of the free
energy, that there is no SDW+SC phase for $s^{++}$ gap even when 
${\delta}_0={\delta}_2$ are not small. 
We analyzed the sign
of $\chi$ for larger ${ \delta}_0$ and ${ \delta}_2$ using our analytical formulas
and confirmed their result.  
In \rfig{fig:GLpp} we show the behavior of
$\chi({ \delta}_0,{ \delta}_2)$ 
at the transition point $T_s({ \delta}_0, { \delta}_2)=T_c$ 
for three representative cases: $\chi({ \delta}_0,0)$,
$\chi(0,{ \delta}_2)$, and $\chi({ \delta}_0,{\delta}_2={ \delta}_0)$. 
In all cases, when $B>0$ 
(e.g., our GL analysis is valid) $\chi({ \delta}_0,{ \delta}_2)$ remains negative.  

We caution, however, that the absence of co-existence between $s^{++}$ SC and
SDW states within GL model does not imply that the two states are always
separated by first-order transition.   GL analysis is only valid near 
$T_s({ \delta}_0, { \delta}_2)=T_c$, when both orders are weak. The situation
at lower $T$ has to be analyzed without expanding in $m$ and ${ \Delta}$.  And,
indeed, we did find a small co-existence region  $T=0$, see \rfig{fig:SSzeroT}.

\subsection{Zero-temperature limit}

We consider only the case of $s^{\pm}$ SC and the limit when relevant
$\delta_0$ and $\delta_2$ are small, i.e., when $T_s/T_c = 1 + \delt$ and
$\delt \ll 1$. We compare energies for pure SDW and SC state and for the
co-existence state and find the region where the co-existence state is
energetically favorable.  

For this,  we first verified that, at small  $\delta_0$ and $\delta_2$, the
values of SDW and SC order parameters at $T=0$ remain the same as at
$\delta_2=\delta_0 =0$, i.e., $m = m_0 = 0.28073 \times (2\pi T_s)$ and $\Delta
= \Delta_0 = m_0 (T_c/T_s)$. These values only change at large enough
$\delta_0$ and $\delta_2$, e.g., $m$ changes when $\delta_0 + \delta_2 >m_0$.    

The free energies of pure SDW and SC states for $m, \Delta > \delta_0 +
\delta_2$ can be straightforwardly evaluated at $T=0$ by replacing the
frequency sums in (\ref{eq:feqc}) by integrals. We obtain
\bea
\cF(m) & = & 
-\frac{m^2}{4}+\frac{\delta_0^2}{2}+\frac{\delta_2^2}{4}+\frac{m^2}{2}\ln\frac{m}{m_0} \\
\cF (\Delta) & = & 
-\frac{\Delta^2}{4}+\frac{\Delta^2}{2}\ln\left(\frac{\Delta}{\Delta_0}\right)
\eea
These free energies have minima at $m=m_0$ and $\Delta=\Delta_0$, respectively. 
 At the minima,
\bea
\cF(m_0) & = & 
-\frac{m^2_0}{4}+\frac{\delta_0^2}{2}+\frac{\delta_2^2}{4} \nonumber \\
\cF (\Delta_0) & = & 
-\frac{\Delta^2_0}{4} = - \left(\frac{T_c}{T_s}\right)^2 \frac{m^2_0}{4}.
\eea
Observe that $\cF(\Delta_0) < \cF(m_0)$ when $T_c=T_s$. This is the consequence
of the fact that SDW magnetism is destroyed by doping and ellipticity, while
superconductivity is unaffected. 

Comparing $\cF(m_0)$ and $\cF (\Delta_0)$, we find that
the  first order transition between pure SDW and SC states occurs at
\be
m^2_0 \delt = \delta_0^2+\delta_2^2/2.  
\label{eq:tzeroT}
\ee
If there is no intermediate co-existence phase,  the SDW state is stable
for  $\delta_0\leq \sqrt{\delt m^2_0 -\delta^2_2/2}$, while 
SC state is stable for larger values of $\delta_0$.

We next determine when the intermediate state appears at $T=0$. For this we
expand the free energy near the SDW and SC states in powers of $\Delta$ and
$m$, respectively. We then obtain, near the SDW state,
\be
\cF (m, \Delta)=\cF (m_0) +a_\Delta \Delta^2+b_\Delta\Delta^4,
\ee
 and near the SC state 
 \be
\cF (m, \Delta)=\cF (\Delta_0) +a_m m^2+b_m  m^4.
\ee
 We verified that $b_\Delta$ and $b_m$ are positive, while $a_m$ and $a_\Delta$
 can be of either sign. The key issue is what are the signs of $a_m$ and
 $a_\Delta$ at the point where  $\cF(m_0) =\cF (\Delta_0)$. We found that, to
 leading order in $\delt$, $a_m = a_\Delta =a$ at this point, and $a$ is
 given by 
\be
a = \frac{\delt^2}{6} \left(1-8z +7z^2 \right),~~z = \frac{\delta^2_2}{2 
m^2_0 \delt}
\label{dr_2}
\ee
Note that to obtain $a$ we had to expand to order $\delt^2$. By virtue of
\req{eq:tzeroT}, $\delta_0 = m_0 \sqrt{\delt}  \sqrt{1-z}$, i.e. we have to
consider  $z \leq 1$.

When $a$ is positive,  both pure states are stable, and there is a first-order
transition between them. When $a<0$,  the pure SDW and SC states are already
unstable at the point where  $\cF(m_0) =\cF (\Delta_0)$, what implies that 
 when we vary $\delta_0$ at a fixed $\delta_2$, there is a range of 
$\delta_0$ around  $\delta_0 = m_0 \sqrt{\delt}  \sqrt{1-z}$ in which
 the co-existence state has a lower energy than the pure states. From
 (\ref{ch_2}) we see that  $a>0$ when $z <1/7$, while $a<0$ for $1/7 <z< 1$. In
 terms of $\delta_2$, this implies that the transition at $T=0$ is first order
 between pure states when $\delta_2 < 0.535 m_0 \sqrt{\delt}$, while at
 larger $\delta_2$, pure SDW and SC phases are separated along $\delta_0$ line
 by the region of the co-existence phase. The width of the co-existence phase
 initially increases as $\delta_2$ increases, but then begins to shrink and
 vanishes when $\delta_2$ approaches $\delta_2 = 1.414 m_0 \sqrt{\delt}$
 ($z$ approaches one from below). At this point, the co-existence region
 shrinks to a point $\delta_0 =0$. 
 At larger $\delta_2$, the SC state has lower energy than the SDW state
 for all values of $\delta_0$ 

If we keep $\delta_0$ fixed but vary $\delta_2$, the co-existence range appears
at $\delta_0 = +0$ ($z=1$) and exists up to $\delta_0 = 0.926 m_0
\sqrt{\delt}$ ($z=1/7$). At larger $\delta_0$ ($z <1/7$), there is a first
order transition between pure SDW and SC states.

\subsection{The phase diagram}  

We now combine the results of GL analysis near the crossing point and at $T=0$
 into the phase diagrams.
For definiteness, we set $\delt = T_s/T_c-1$ to be small and consider the set
of phase diagrams in variables $T$ and $\delta_0$ for different fixed
$\delta_2$. The results of this subsection has to be
 compared with the phase diagrams presented in panels (a1)-(a3) in
 Fig. \ref{fig:tsc2pd}, see also Fig.~\ref{fig:pdSPzeroT}. 

From the analysis in the preceding two subsections, we found five critical 
 values of $\delta_2$: two are obtained from the GL analysis of the range of
 the co-existence phase, and are given by 
 (\ref{dr_5}), two are critical values at which the co-existence phase first
 appears and then disappears at $T=0$, and the last one is the maximum value of
 $\delta_2$ at which $T_s (\delta_0 =0, \delta_2) = T_c$. From (\ref{dr_1})
 this value is $\delta_2 =1.739 m_0 \sqrt{\delt}$. Arranging these five
 values from the smallest to the largest, we obtain the following set of phase
 diagrams at
 small $\delt$: 
\begin{itemize}
\item[(a)] 
For $\delta_2 < 0.535 m_0 \sqrt{\delt}$, there is no intermediate phase, and
pure SDW and SC transitions are separated by a line of a fist-order transition.
The line is tilted towards smaller $\delta_0$ at smaller $T$: it 
originates at $\delta_0 = m_0 \sqrt{\delt} 
(1.508 - 0.5 \delta^2_2/(m^2_0 \delt))^{1/2}$ at $T=T_c$ and ends up at
$\delta_0 = m_0 \sqrt{\delt} (1 - 0.5 \delta^2_2/(m^2_0 \delt))^{1/2}$. 
\item[(b)]
For  $0.535 m_0 \sqrt{\delt} < \delta_2 < 0.826 m_0 \sqrt{\delt} <$, 
the intermediate phase appears near $T=0$ and extends to some $T <T_c$. At
larger $T$, the transition remains first order.  This behavior is consistent
with the panel (a1) in Fig. \ref{fig:tsc2pd}  
\item[(c)]
For  $0.826 m_0 \sqrt{\delt} <\delta_2 < 1.414 m_0 \sqrt{\delt}$, 
the intermediate phase occupies the whole region $T <T_c$.
 This behavior is consistent with the panel (a2) in Fig. \ref{fig:tsc2pd}  
\item[(d)]
For  $1.414 m_0 \sqrt{\delt} <\delta_2 < 1.663 m_0 \sqrt{\delt}$, 
SC state wins at $T=0$ for all $\delta_0$. There is phase transition at finite
$T$. The transition is first order between pure SDW and SC state at smaller
$T$, but the co-existence phase still survives near $T_c$.   This behavior is
consistent with the panel (a3) in Fig. \ref{fig:tsc2pd}     
\item[(e)]
For  $1.663 m_0 \sqrt{\delt} <\delta_2 < 1.739 m_0 \sqrt{\delt}$,
the co-existence phase near $T_c$ disappears, and the  transition becomes 
first-order along the whole line separating SDW and SC states.
\item[(f)]    
$\delta_2 > 1.739 m_0 \sqrt{\delt}$, $T_s (\delta_0,\delta_2)$ becomes
smaller than $T_c$ for all $\delta_0$, and the system only develops a SC order.
\end{itemize}
This behavior is also totally consistent with Fig.~\ref{fig:pdSPzeroT}: 
all different phase diagrams are reproduced if we take horizontal cuts at
different $\delta_2$. We see therefore that numerical and analytical analysis
is in full agreement with each other.

\begin{figure}[t]
\centerline{\includegraphics[width=0.5\linewidth]{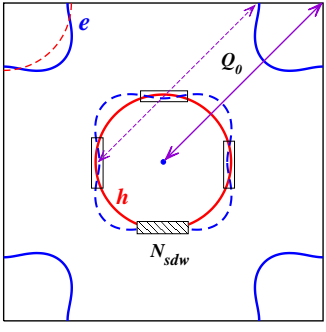}}
\caption{(color online) 
	A partially-gapped SDW state, where only a fraction $N_{sdw}$ of 
	the electron and hole FSs is nested. On the remaining parts 
	the dispersions  $\xi_f$ and $\xi_c$  are very different. 
	The SDW state appearing below $T_s$ gaps excitations only in 
	the shaded/boxed areas (see \rfig{fig:excitations}(d)),  
	while on the rest of the FSs the  dispersions are close to the original
	$\xi_f$ and $\xi_c$. The SC state below below $T_c$ does not compete with
	SDW in non-nested regions, but  competes with SDW state in the nested
	(boxed) regions. 
}
\label{fig:NsdwFS}
\end{figure}

The only result of numerical studies not reproduced in small $\delt$
analytical expansion is the existence of a range of $\delta_2$ where the
transition between the SDW phase and the co-existence phase is second order,
while the transition between the SC phase and the co-existence phase is first
order, see Fig. \ref{fig:pdSPzeroT}. To reproduce this effect in analytical
treatment, we would have to expand to the next order in $\delt$. Note in
this regard that it is evident from Fig.  \ref{fig:pdSPzeroT} that the width of
the range where one transition is first order and another is second order
shrinks as $T_s/T_c$ decreases.    

\section{\label{sec:Nsdw} Partial SDW state }
\label{sec:partSDW}

In previous Sections we considered the situation when the splitting between
hole and electron FSs is small.  We now consider how the phase diagram is
modified if in some $k$-regions hole and electron FSs are quite apart from each
other (after we shift the hole FS by $\vQ_0$). Such regions are far from
nesting and we make a simple assumption that they are not affected by SDW. We
then split the FS into nested parts where commensurate SDW state exists and a
SC order can exist as well, 
and non-nested parts, where only SC order is possible. We present this
schematically in \rfig{fig:NsdwFS}. 
The nested parts lie in some intervals of angles $\phi$ with total
circumference $\Del \phi$, and have weight $N_{sdw} < N_{total} = 1$
($\Del\phi/2\pi = N_{sdw}/N_{total}$). 

The free energy and the self-consistency equations then can be 
written as  sums of the two 
contributions. The first sum is over the FS part that has only 
SC order parameter, and in the second sum we integrate over part of the 
FS with both orders.~\cite{Machida81sdwsc} 
\begin{widetext}
\bea
&& \frac{\Del F(\Delta, m)}{4N_F} = 
\frac{\Delta^2}{2} \ln\frac{T}{T_c} + N_{sdw} \frac{ m^2}{2} \ln\frac{T}{T_s} 
 - 2\pi T \sum_{\vare_n>0} 
(1-N_{sdw}) \left[ \sqrt{\vare_n^2+\Delta^2} - |\vare_n| - \frac{\Delta^2}{2|\vare_n|} \right] 
\\ \nonumber 
&& \hspace*{6cm} - 2\pi T \sum_{\vare_n>0} \int\limits_{\Del\phi} \frac{d\phi}{2\pi} 
Re \left[ \onehalf(\Sigma_+ + \Sigma_-)
- |\vare_n| - \frac{\Delta^2}{2|\vare_n|} - \frac{m^2}{2|\vare_n|}
\right] \,, 
\\
&& \Delta \ln\frac{T}{T_c} = 
2\pi T \sum_{\vare_n>0} 
(1-N_{sdw}) \left[ \frac{\Delta}{\sqrt{\vare_n^2+\Delta^2}} - \frac{\Delta}{|\vare_n|} \right]
+ 2\pi T \sum_{\vare_n>0} \int\limits_{\Del\phi} \frac{d\phi}{2\pi} 
Re \left[ \onehalf\left( \pder{\Sigma_+}{\Delta} + \pder{\Sigma_-}{\Delta}\right)
- \frac{\Delta}{|\vare_n|} \right] \,,
\label{wed_1}
\\
&& N_{sdw} m \ln\frac{T}{T_s} = 
2\pi T \sum_{\vare_n>0} \int\limits_{\Del\phi} \frac{d\phi}{2\pi} 
Re \left[ \onehalf\left( \pder{\Sigma_+}{m} + \pder{\Sigma_-}{m}\right)
- \frac{m}{|\vare_n|} \right] \,. 
\label{wed_2}
\eea
\end{widetext}

The self-consistency equations (\ref{wed_1}) and (\ref{wed_2}) 
are obtained by  minimization of the functional $\Del F$, 
$\partial (\Del F) / \partial \Delta = 0$ and $\partial F / \partial m = 0$, 
and these expressions reduce to previous 
formulas (\ref{eq:scSCqc})-(\ref{eq:feqc}) for $N_{sdw}=1$. 

\begin{figure}[t]
\centerline{\includegraphics[width=\linewidth]{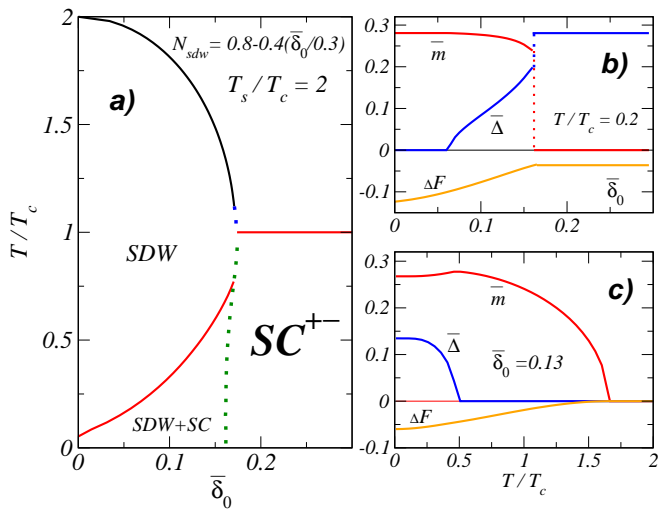}}
\caption{(color online) 
	(a) The phase diagram for $s^\es$ superconductivity, and SDW order
	parameter existing only in boxed regions of the FS in \rfig{fig:NsdwFS},
	with the relative width $N_{sdw}$.  We manually set  the doping
	dependence of $N_{sdw}$ to be $N_{sdw} = 0.8 - 0.4 ({\bar\delta}_0 / 0.3)$
	and  neglected the effect of this variation of $N_{sdw}$ on $T_s$.
	Observe that SC and SDW orders co-exist in a wide range of ${\delta}_0$ 
	and $T$.  
	(b,c) the order parameters,  $\Delta$ and $m$, and
	the free energy, $F$, as functions of ${\delta}_0$  at a constant
	temperature, $T/T_c=0.2$ (b), and as functions of $T$ at a constant
	${\bar \delta}_0 =0.13$ (c). As a function of ${\delta}_0$, SDW
	order parameter  starts decreasing when SC order appears, and then jumps
	to zero and the system becomes a pure SC.
}
\label{fig:NsdwSPm}
\end{figure}

We find that the results are very similar to what we found within 
the approximation of a small FS splitting. 
The typical picture is shown in \rfig{fig:NsdwSPm}.
  
The  only differences from \rfig{fig:ellipscx} in this case are 
the co-existence of SC and SDW states already at zero doping $\delta_0=0$, 
and weak first order transition to purely SC state. 
We also analyzed $s^{++}$ SC order and again found a 
much weaker tendency for co-existence, similar to \rfig{fig:SSzeroT}.

\section{Conclusions}
\label{sec:concls}

To conclude, we presented a general theoretical description of
the interplay between itinerant SDW and SC orders in two-band metals.  Within
the mean-field approach we derived coupled self-consistency equations for the
order parameters and the expression for the free energy, which is necessary to
determine the stability of different phases. 

We considered the FS geometry  with one hole and one electron bands of
different shapes (a simplified FS geometry for Fe-pnictides) and investigated
the phase diagrams and the stability of the SDW+SC states for: 
(a) different gap structures of the SC state, 
Figs.~\ref{fig:ellipscx}, \ref{fig:pdSPzeroT} vs. \ref{fig:SSzeroT}; 
(b) variations in the relative strength of SDW and SC interactions, 
Figs.~\ref{fig:pdSPzeroT}, \ref{fig:cx35}; 
(c) ellipticity of electron pockets, 
Figs.~\ref{fig:tsc2pd}, \ref{fig:ellipscx}, \ref{fig:pdSPzeroT}; and 
(d) incommensuration of SDW order,  
Figs.~\ref{fig:cx35}, \ref{fig:FFLO}. 
We considered the case when the transition temperature to pure
SDW state, $T_s$, is higher than the
critical temperature $T_c$ of a pure SC state. In the opposite case, $T_s<T_c$,
the SC state develops first and suppresses SDW state. 

We found that the SC $s^\pm$ state with extended $s$-wave symmetry has much
stronger affinity with the SDW state than the traditional $s^\pp$ state. 
A co-existence region of $s^\pp$ SC state with SDW is 
tiny and the co-existence is anyway very weak in terms of energy gain compared to
the pure SDW state. The transition from the pure SDW state to the pure SC state
is always first order, \rfig{fig:SSzeroT}. 
For  $s^\pm$ gap, there is a stronger inclination towards
co-existence with SDW state due to effective ``attraction'' between the two
orders. We found that, depending on the interplay between different effects
(e.g., ellipticity and doping), the transition between SDW and SC orders is
either first order or continuous, via the intermediate SDW+SC phase, in which
both order parameters are non-zero, 
Figs.~\ref{fig:pdSPnocx}-\ref{fig:ellipscx}.\cite{d_footnote}

We note  that the co-existence region gets larger with increased strength of the
SDW interaction relative to its SC counterpart, described by the ratio 
$T_s/ T_c$. Thus generally we should see better co-existence between SDW and SC
states, if $T_s$ is increasingly larger than $T_c$, \rfig{fig:pdSPzeroT}. 

Our results are in a disagreement with a common belief that, because SDW and
SC states compete for the Fermi surface, the SDW+SC state should emerge when
a pure SDW state next to the boundary of the co-existence region still has a
modified Fermi surface at $T=0$, and should not emerge when  fermionic
excitations in the pure SDW phase are fully gapped at zero temperature.  We
found that the key reason for the existence of the mixed SDW+SC state is the
``effective attraction'' between the SDW and SC orders, while the presence or
absence of the Fermi surface in the SDW state at $T=0$ matters less.
Specifically,  we found cases when SDW and SC orders do co-exist even when
fermionic excitations in the pure SDW phase are fully 
gapped at $T=0$, \rfig{fig:pdSPzeroT}(a), 
and we also found, for $s^{++}$ pairing, that there might be no co-existence 
down to $T=0$ even when  the pure SDW phase has a Fermi surface, \rfig{fig:SSzeroT}.

The phase diagrams for $s^{+-}$ gap  are quite consistent with the experimental
findings in pnictides.  For example, first order transition in
\rfig{fig:pdSPnocx} looks very similar to phase diagram of 1111 materials
(La,Sm)OFeAs, where FSs are more cylindrical.  The co-existence region in
\rfig{fig:ellipscx} correlates well with doped 122 materials based on BaFe$_2$As$_2$,
where hole and electron FSs are less nested.  And \rfig{fig:pdSPzeroT} shows
that one can  get both SDW+SC phase and first-order transitions for the same SC
state and the same family of materials.  Our key result is that the way the
doping is introduced into the sample will determine  the nature of the FS
changes, and the path it will take in the $(\delta_0, \delta_2)$-plane:
whether through a first order transition or through a co-existence region.  In
other words, we argue that there is strong correlation between how exactly FSs
evolve upon doping and  whether or not SC and SDW states co-exist.

The final remark.  
In the literature, there exists a notion of 
``homogeneous'' and ``inhomogeneous'' co-existence of SC and SDW orders. 
The latter is a metastable state when the two orders 
exist in \emph{different} spatial parts of the material.
What we emphasize is that the other kind, ``homogeneous'' co-existence 
of SC and SDW orders in real space, is in fact ``inhomogeneous'' 
in momentum space: the SC and SDW orders dominate excitation gaps on 
\emph{different} parts of the FS. 

\subsection{Acknowledgment}

We acknowledge with thanks useful discussions with 
D.~Agterberg, V.~Cvetkovic, R.~Fernandes, I.~Eremin, 
I.~Mazin,  J.~Schmalian, O.~Sushkov, and Z.~Tesanovic.
A.V.C. acknowledges the support from NSF-DMR 0906953. \\


\appendix

\section{\label{appA} Electron and Hole dispersion for small FS splitting}

In this Appendix, we discuss in detail the approximation we used 
 for the dispersions of holes and electrons for the case when the splitting
between hole and electron FSs is small.

SDW and SC orders mix $c$-fermions with momenta $\vk$ and $f$-fermions with 
 momenta $\vk + \vq$. The generic expressions for the two dispersions are 
\be
\xi_c (\vk) = \mu_c - \frac{(\vk)^2}{2m_c}, ~~ \xi_f (\vk +\vq) = 
\frac{(k+q)_x^2}{2m_{fx}} + \frac{(k+q)_y^2}{2m_{fy}} -\mu_f 
\ee
When the two FSs are circles of non-equal size, $m_{fx} = m_{fy} = m_f$, we
have  $(\mu_c,m_c) \approx (\mu_f, m_f) \approx (\mu, m)$, but $m_c \neq m_f$
and $\mu_c \neq \mu_f$. The approximation we used in the text implies that  
 \begin{widetext}
 \bea
 &&\xi_c (\vk) = \mu_c - \frac{(k+q)^2}{2m_c}  = \frac{\mu_c + \mu_f}{2}
 -\frac{k^2}{4} \left(\frac{1}{m_c} + \frac{1}{m_f}\right)- \frac{\vk \vq}{4}
 \left(\frac{1}{m_c} + \frac{1}{m_f}\right) 
 \nonumber \\
 &&+ \frac{\mu_c - \mu_f}{2} -\frac{k^2}{4} \left(\frac{1}{m_c} -
 \frac{1}{m_f}\right)+ \frac{\vk \vq}{4} \left(\frac{1}{m_c} +
 \frac{1}{m_f}\right) 
 \nonumber \\
 && \approx \frac{\mu_c + \mu_f}{2} -\frac{k^2}{4} \left(\frac{1}{m_c} +
 \frac{1}{m_f}\right) - \frac{\vk \vq}{4} \left(\frac{1}{m_c} +
 \frac{1}{m_f}\right)
 \nonumber \\
 && + \frac{\mu_c - \mu_f}{2} + \frac{k^2_F}{4m} \left({m_c} - {m_f}\right) + 
 \frac{{\vk}_F \vq}{2m}
 \nonumber \\
 &&\xi_f (\vk +\vq) = \frac{(k+q)^2}{2m_f} -\mu_f  =  \frac{k^2}{4}
 \left(\frac{1}{m_c} + \frac{1}{m_f}\right) + \frac{\vk \vq}{4}
 \left(\frac{1}{m_c} + \frac{1}{m_f}\right) -\frac{\mu_c + \mu_f}{2}  
 \nonumber \\
 && + \frac{\mu_c - \mu_f}{2} -\frac{k^2}{4} \left(\frac{1}{m_c} -
 \frac{1}{m_f}\right)+ \frac{\vk \vq}{4} \left(\frac{3}{m_f} -
 \frac{1}{m_c}\right) 
 \nonumber \\
 && \approx  \frac{k^2}{4} \left(\frac{1}{m_c} + \frac{1}{m_f}\right) 
 + \frac{\vk \vq}{4} \left(\frac{1}{m_c} + \frac{1}{m_f}\right) - 
 \frac{\mu_c + \mu_f}{2} 
 \nonumber \\ 
 && + \frac{\mu_c - \mu_f}{2} + \frac{k^2_F}{4m} \left({m_c} - {m_f}\right) + 
 \frac{{\vk}_F \vq}{2m}
 \label{nn_2}
 \eea
 Introducing $\xi_{\vk \vq}$ and $\delta_{\hvk \vq}$ defined in (\ref{nnnn_1}), we obtain
 \bea
 \xi_{\vk \vq} &=&  \frac{k^2}{4} \left( \frac{1}{m_c} + \frac{1}{m_f}\right) 
 + \frac{\vk \vq}{4} \left(\frac{1}{m_c} + \frac{1}{m_f}\right) - 
 \frac{\mu_c + \mu_f}{2}  \nonumber \\
 \delta_{\hvk \vq} &=&  \frac{\mu_c - \mu_f}{2} + \frac{k^2_F}{4m} \left({m_c} - {m_f}\right) + 
 \frac{{\vk}_F \vq}{2m} = \frac{1}{2} \vv_F \left(\vk^c_F - \vk^f_F - \vq \right) 
 \label{nn_3}
 \eea   
 \end{widetext}
We emphasize that, within this approximation,  $\xi_{\vk \vq}$ and
$\delta_{\hvk \vq}$ are two independent variables, one depends on the deviation
along the FS in the transverse direction, another depends on the angle along
the FS. 
When $m_{fx} \neq m_{fy}$, the derivation remains the same but 
$\delta_{\hvk \vq}$ acquires  an  additional term with the difference between
$m_{fx}$ and $m_{fy}$ (the $\cos 2 \phi$ term).


\begin{thebibliography}{10}
\bibitem{d_footnote}
{A $d$-wave symmetry gap presumably lies in
between these two cases as some parts of the nested hole and electron FSs will
have the gaps of the same sign, while other parts will have the opposite
signs.\cite{Fernandes09}}
\end{thebibliography}

\end{document}